\documentclass[10pt,conference]{IEEEtran}
\IEEEoverridecommandlockouts

\usepackage{graphicx} % Required for inserting images
\usepackage{subcaption}
\usepackage{tabularx}
\usepackage{algorithm,algorithmic}
\usepackage{soul}
\usepackage{multirow}
\usepackage{mathtools} 
\usepackage{amsmath} 
\usepackage{bbding}
\usepackage{hyperref}
\usepackage{cite}
\usepackage{amsmath,amssymb,amsfonts}
\usepackage{algorithmic}
\usepackage{graphicx}
\usepackage{textcomp}
\usepackage{xcolor}

\usepackage{graphicx}
\usepackage{subcaption}
\usepackage{tabularx}
\usepackage{algorithm,algorithmic}
\usepackage{soul}
\usepackage{multirow}
\usepackage{url}
\usepackage{hyperref}
\usepackage{color}
\usepackage{color}
\usepackage{cite}
\usepackage{amsmath,amssymb,amsfonts}
\usepackage{algorithmic}
\usepackage{graphicx}
\usepackage{textcomp}
\usepackage{xcolor}

\usepackage{graphicx}
\usepackage{subcaption}
\usepackage{tabularx}
\usepackage{algorithm,algorithmic}
\usepackage{soul}
\usepackage{multirow}
\usepackage{url}
\usepackage{hyperref}

\usepackage{microtype}
\usepackage{balance}

\usepackage{color}

\let\emptyset\varnothing
\newcommand{\tool}{\textsc{GARL}}

\author{
\IEEEauthorblockN{
Linfeng Liang\textsuperscript{1}, Yao Deng\textsuperscript{1}, Kye Morton\textsuperscript{2}, Valtteri Kallinen\textsuperscript{2}, 
Alice James\textsuperscript{1}, Avishkar Seth\textsuperscript{1},
}
\IEEEauthorblockN{
Endrowednes Kuantama\textsuperscript{1}, Subhas Mukhopadhyay\textsuperscript{1}, Richard Han\textsuperscript{1}, Xi Zheng\textsuperscript{1,*}
}
\IEEEauthorblockA{\textsuperscript{1}\textit{School of Computing, Macquarie University, Australia}} 
\IEEEauthorblockA{\textsuperscript{2}\textit{Skyy Network, Australia}} 
\thanks{\textsuperscript{*}Corresponding author. Email: james.zheng@mq.edu.au}
}

\begin{document}
% \vspace*{0.1cm}
%%
%% The "title" command has an optional parameter,
%% allowing the author to define a "short title" to be used in page headers.
% \title{RLaGA: A Reinforcement Learning Augmented Genetic Algorithm For Generating Real and Diverse Marker-Based Landing Violations Dynamically}

\title{\tool: Genetic Algorithm-Augmented Reinforcement Learning to Detect Violations in Marker-Based Autonomous Landing Systems}
\maketitle
\begin{abstract}

Automated Uncrewed Aerial Vehicle (UAV) landing is crucial for autonomous UAV services such as monitoring, surveying, and package delivery. It involves detecting landing targets, perceiving obstacles, planning collision-free paths, and controlling UAV movements for safe landing. Failures can lead to significant losses, necessitating rigorous simulation-based testing for safety. Traditional offline testing methods, limited to static environments and predefined trajectories, may miss violation cases caused by dynamic objects like people and animals. Conversely, online testing methods require extensive training time, which is impractical with limited budgets. To address these issues, we introduce \tool, a framework combining a genetic algorithm (GA) and reinforcement learning (RL) for efficient generation of diverse and real landing system failures within a practical budget. \tool \ employs GA for exploring various environment setups offline, reducing the complexity of RL's online testing in simulating challenging landing scenarios. Our approach outperforms existing methods by up to 18.35\% in violation rate and 58\% in diversity metric. We validate most discovered violation types with real-world UAV tests, pioneering the integration of offline and online testing strategies for autonomous systems. This method opens new research directions for online testing, with our code and supplementary material available at \url{https://github.com/lfeng0722/drone_testing/}.
\end{abstract}

\maketitle       

%\section{Introduction}

% \section{Background}
% \label{bg}

% \subsection{Genetic Algorithm}

% \subsection{Reinforcement Learning}
% \par 

% \sectionspacing
\section{Introduction}

Automated UAV landing, such as the routines available in flight controllers like Ardupilot~\cite{ardupilot}, is vital during final flight stages for targeting ground landing spots. With the growth of autonomous UAV services—monitoring, surveying, and delivery~\cite{vetrella2015cooperative, shakhatreh2019unmanned, mittal2019vision}—the reliance on non-manually controlled UAVs is increasing. The addition of ground-truth markers can significantly reduce the risk of accidents \cite{baca2019autonomous, lee2023intelligent, marcu2018safeuav}. Marker-based landing systems are widely used in logistical and industrial UAV applications to ensure reliable landings for large or expensive payloads \cite{Skyy_Windjammer_Landing_Resort, Skyy_MedLife_Skyy_Network, Swoop_Aero_Kite}. Additionally, “drone in a box” solutions, which require precision landings (less than 10 cm) to charging or docking stations, are an emerging sector needing this capability \cite{DJI_Enterprise_DJI_Dock2, Dronehub, Percepto_Drone_in_a_Box}. However, automated marker-based landings carry risks: failed landings due to adverse weather, incorrect marker identification, and collisions with objects. Given stringent safety regulations from authorities like the Federal Aviation Administration (FAA) or European equivalent, and recent incidents, there's a highlighted need for thorough testing of automated marker-based landings to ensure compliance and safety~\cite{news2, news1, accidentnews,chen2020robust}.

\par 
% The test of marker-based landing is similar to the test of Autonomous driving systems (ADSs). According to recent research \cite{li2020av, deng2022scenario}, the test of ADSs can be regarded as the test of a simulation environment and an AI-based cyber-physical system. For example, in the test of ADSs, much research \cite{li2020av, deng2022scenario, haq2022many} was conducted by using LGSVL \cite{rong2020lgsvl} as the simulation tool and Apollo \cite{ap} as the AI-based cyber-physical system. 
% Differently, the marker-based landing is not as tightly regulated as the driving task i.e. the only violation in marker-based landing is the UAV does not land on the right marker. 
Simulation testing is commonly used for Autonomous Driving Systems (ADSs) \cite{li2020av, deng2022scenario, haq2022many, lou2022testing}. UAV operations, including marker-based landings, can similarly be tested using simulation environments like \textit{AirSim} \cite{AirSim2017fsr}. Recent studies highlight the potential of offline testing for AI-based cyber-physical systems \cite{tian2022mosat}. However, offline approaches like GA rely on pre-defined configurations for variables such as weather and object positions, limiting their ability to explore the dynamic search space and potentially missing critical corner cases. In contrast, online methods like RL can adjust test cases in real-time but often struggle to converge within limited time due to the extensive learning space in simulation testing \cite{feng2023dense}.

\begin{figure}
    \centering
    \includegraphics[width=1\linewidth]{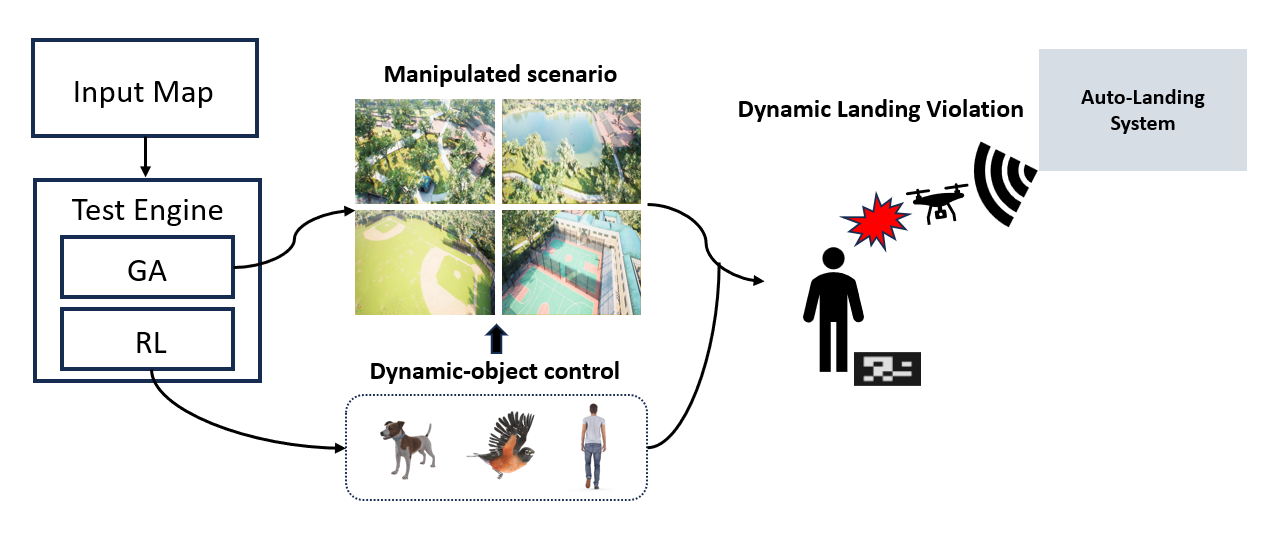}
	\caption{A high-level overview of \tool \ framework, including the simulator and the test engine. }
\label{overall}  
\end{figure}
\par A new method, GA-augmented RL (\tool), is introduced to address the described challenges.  Figure \ref{overall} depicts a high-level overview of \tool. This approach utilises GA for offline exploration of environmental setup and RL for online management of dynamic object's movements. As such, GA creates varied environmental conditions in the testing map, while RL manipulates dynamic objects to interact with the auto-landing UAV.
This hybrid strategy offers an efficient testing solution, capturing a wide array of violations within time constraints, due to the reduction in complexity for RL. Our experiments validate the effectiveness of this integrated approach in overcoming the identified challenges. \tool\ increases the violation rate by up to 18.35\% and increases the diversity of generated violation cases by more than 58\% compared to the current state-of-the-art (SOTA) methods. Moreover, most of these additional violation cases have been confirmed as potential violations in real-world settings. Our contributions can be summarised as follows:

\begin{itemize}
    \item A novel integrated GA-augmented RL (\tool) method to efficiently generate diverse and realistic corner test cases for marker-based UAV landing systems. 
    %We introduce a customized offline GA using our chromosome representation and variation operations to increase the diversity of found corner cases. While we also introduce online RL to manipulate the dynamic-
    
    %with two types of genes to control conditions such as weather and RL-guided online manipulation dynamic-objects in the scene.  %\jz{this contribution seems very thin (RICK: this comment was directed to the old second contribution, but I thought to leave it here in case we can use it to improve the contribution wording), I would suggest to highlight our contribution of customized GA using our chromesome representation with two types of genes and our design of DRL}.
    \item An extensive demonstration of simulation experiments showing that \tool \ works across different landing systems, increases violation rate by up to 18.35\%, and increases the diversity of generated violation cases by more than 58\% in comparison to state-of-the-art baselines in various settings.
    
    % \jz{I would suggest adding some good statistics from our evaluation here and highlight we compare with sota baselines. Also add a sentence to give our repository link which needs to be anonymous for code and experiment data.}
    \item A demonstration of equivalent real-world UAV flight tests that validate various simulation-detected violations in recreated practical environments.
    
    %We conducted real-world tests and confirmed most types of violations found in the simulation by flying the UAV in practical environments while running automated landing software. 
    
    % We confirm that we are able to reproduce in the real-world all types of the landing violations found in simulation by flying drones in practical environments while running automated landing software.
    
%    \item \textcolor{red}{Rick: REMOVE?  We propose different metrics to evaluate the diversity of generated test cases for UAV landings}\jz{this contribution seems very thin, I would suggest to highlight our contribution of customized GA using our chromesome representation with two types of genes and our design of DRL}.
%    \item \textcolor{red}{Rick: REMOVE?  We conduct extensive experiments to verify our proposed method in both simulation and the real-world.} \jz{I would suggest adding some good statistics from our evaluation here and highlight we compare with sota baselines. Also add a sentence to give our repository link which needs to be anonymous for code and experiment data.}
\end{itemize}
% This paper is organized as follows: In section \ref{RW}, we provide a review and discussion of the related work relevant to our study. Section \ref{method} details how we created our search method. In section \ref{exp}, we indicate the design of our experiments. In section \ref{result1}, we present the results of our comprehensive experiments based on the methodology described in section \ref{method}. In section \ref{TV}, we discuss potential threats against our proposed methods.  Conclusions are summarized in section \ref{conclusion}.
% \aftersectionskip

% \sectionspacing
\section{Related Work}
% \vspace{-0.2cm}
\label{RW}
% The GA mimics biological evolution. In the beginning, the GA generates the initial population, and a fitness function is defined which could be understood as an optimisation objective. Then every single individual in the initial population through continuous mutation and crossover between each other achieves a higher fitness value.
% GA has the ability to generate corner cases from the initialization \cite{}
% \subsectionspacing

\subsection{GA-based Test Generation}
\par The efficacy of GA has been demonstrated in searching for corner test cases that cause violations and failures in AI-based cyber-physical systems \cite{li2020av, schmidt2022stellauav,panichella2015reformulating,abdessalem2018testing,ben2016testing,ebadi2021efficient,huai2023sceno,luo2021targeting}. GA has extensive applications in search-based software testing, and is usually employed as an offline search technique \cite{wegener2004evaluation}.
In~\cite{abdessalem2018testing,ben2016testing}, the authors proposed new objective functions containing multiple test objectives to guide the search of test scenarios for evaluating Advanced Driver Assistance Systems (ADAS). In~\cite{ebadi2021efficient}, a search-based method was proposed to test the pedestrian detection algorithm in Baidu Apollo~\cite{Apollo} by manipulating static parameters such as weather and the initial positions of dynamic objects in the test scenarios. AV-FUZZER \cite{li2020av} employs both a global fuzzer and a local fuzzer based on GA to search for corner cases in ADSs. StellaUAV \cite{schmidt2022stellauav} uses existing GA and different optimisation methods to search for corner cases specifically for obstacle avoidance components. AutoFuzz \cite{zhong2022neural} demonstrates that neural networks have the potential to augment GA and employ a gradient as an indicator to mutate seeds. ScenoRITA~\cite{huai2023sceno} proposed new gene representations for testing scenarios to make obstacles fully mutable for finding more meaningful violation scenarios.
Recent search-based testing work MOSAT \cite{tian2022mosat} employs GA to manipulate pre-defined driving maneuvers to create diverse violations for ADSs. 

These works focus on new scenario representations, objective functions, and innovative crossover and mutation operations, with test scenarios determined offline. In contrast, we introduce a method that combines a specially designed GA with targeted mutations and crossover operations to enhance the generation of diverse online search spaces for RL. Our approach uses RL to dynamically alter the complex interactions and trajectories of dynamic objects.
%\lf{These related work contributions primarily focus on new scenario representations, new objective functions, and innovative crossover and mutation operations. Test scenarios are determined offline before the execution. Compared to these works, we introduce an innovative method combining a specially designed GA—with targeted mutations and crossover operations to better generate diverse online search space for RL. Our approach involves applying RL to dynamically alter the complex interplay and trajectories of dynamic objects.} 
This strategy aims to increase violation diversity in marker-based landing scenarios across all components of an automatic landing. In our experiments, we conduct a fair comparison between \tool\ and a multi-objective GA method, showcasing superior results.
% \aftersubsectionskip

% \subsectionspacing
\subsection{RL-based Test Generation}
\par The validity of RL has been demonstrated in generating violation cases in software tests \cite{lu2022rgchaser, koren2018adaptive}. RL is a Markov Decision Process (MDP) which is built based on an interactive environment that includes agents, actions, policies and rewards \cite{sutton2018reinforcement}. In the MDP, the agent perceives the current state and performs actions in the environment based on the policy to receive the reward. RL agents can serve as offline fuzzers, altering test scenarios based on a reward function, with studies proving their effectiveness in fuzzing \cite{bottinger2018deep, lu2022rgchaser}. Furthermore, RL can be used for online testing to dynamically create corner cases via a predefined reward function \cite{koren2018adaptive,lu2022learning}. Research shows that RL-based online testing approaches are able to control dynamic objects in the test scenario in real time to explore more possible violation scenarios \cite{koren2018adaptive,lu2022learning}. Additionally, D2RL introduces an online RL-based method to fully comprehend the ADS testing environment, despite its high computational demands and long training periods \cite{feng2023dense}. 

A common challenge for RL-based methods in simulation testing is the restricted budget, impacting RL agent convergence. A backward training strategy \cite{koren2021finding} has been proposed to mitigate this by using expert demonstrations in low-fidelity simulations to train the agent before deployment in high-fidelity simulations, which can cause transferability issues. Unlike existing RL-based methods, our framework uniquely combines GA and RL, using GA to simplify RL's learning dimensions. Our surrogate training eliminates the need for expert demonstrations, enhancing agent transferability and allowing direct application in high-fidelity simulations.
%\lf{A common challenge for these RL-based methods in simulation testing is the restricted budget, impacting RL agent convergence. A backward training strategy has been proposed to reduce the budget required \cite{koren2021finding}. Their method involves acquiring expert demonstrations in low-fidelity simulations to train the agent and then deployed in high-fidelity simulations, which may cause agent transferability problems. Different from existing RL-based test generation methods, our framework uniquely combines GA and RL, using GA to simplify RL's learning dimensions. Our surrogate training regime omits the need for expert demonstrations and enhances the agent's transferability, allowing direct application in high-fidelity simulations.}
% \aftersubsectionskip

% \subsectionspacing
\subsection{Combining GA and RL}
Combining GA and RL, known as Evolutionary Reinforcement Learning (ERL), has been explored in recent years \cite{khadka2018evolution}. ERL utilizes GA to evolve RL policies for addressing path planning and navigation problems \cite{marchesini2020genetic,prathiba2021hybrid,stafylopatis1998autonomous,xu2024deep}. These methods use GA to mutate and evolve RL agent parameters and architectures, focusing on training rather than test generation. Few works combine GA and RL for test generation. For instance, Wuji \cite{zheng2019wuji} employs ERL for game testing to find corner cases revealing logical bugs or crashes. In \cite{esnaashari2021automation}, GA and RL are combined to test traditional software, using Q-learning to explore uncovered paths. However, these methods are not directly applicable to autonomous systems with ML modules due to the lack of conventional control structures. To our knowledge, we are the first to combine GA and RL for the test generation of autonomous landing systems.
\section{Method}
\label{method}

\begin{figure}
    \centering
    \includegraphics[width=0.95\linewidth]{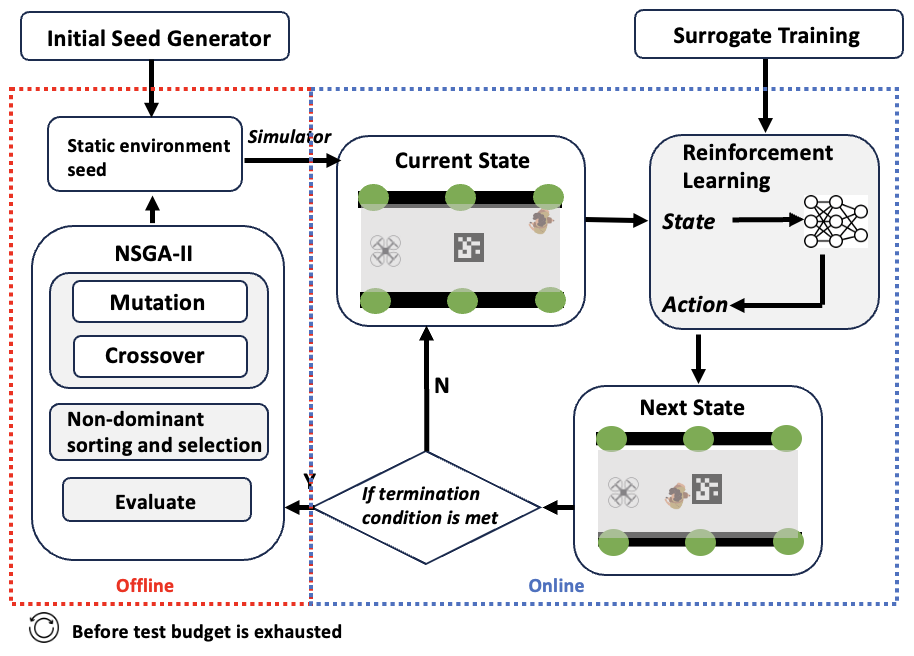}
	\caption{Workflow of \tool}
\label{workflow}  
% \vspace{-0.8cm}
\end{figure}
% The search for violation cases starts with modeling the marker-based UAV auto-landing scenario with a chromosome representation. Generally, 
% \subsectionspacing
\subsection{Overview}
% The aim of \tool \ is to generate real and diverse corner cases in marker-based auto-landing scenarios. Specifiaclly, the scenario can be categories into two main parts: environment configuration and dynamic objects. Environment configuration is searched offline by GA, which includes factors like weather parameters, the marker's position and the NPC-object's information. These configurations remain constant throughout each run and are generated at the beginning of each run. NPC-objects are controlled online by the RL, and the RL agent will keep manipulating the NPC-objects' trajectories throughout each run. 
%\jz{I would rewrite this paragraph, you can start with GA and mention what search space GA needs to explore (environment and the position of the marker) and then talk about RL and mention what search space RL needs to explore (the trajectories of those dynamic objects in the deployed envrionment). Then you can set a scene, our approach is motivated by the real-world deployment environment. We want to set a similar simuation environment as the deployment environment where we will first setup a simulation map which looks like the physical deployment location, use GA to change the environment and the varierty possible places of placing the marker (offline). And then for each of this setting (seed), we want to use trained RL to explore worst case scenario where the input seed can lead to UAV landing violation and start the rest of GA opeartions such as cross over and mutation to find diverse errors guided by our multi-object fitness function.}

\par %Following this, we present an offline GA that is conservative in nature to explore diverse violation seeds. Subsequently, we propose an aggressive online RL algorithm that is designed to explore domain-expert-specified violation cases with high fidelity, while the RL can also benefit the GA. 
The aim of \tool \ is to generate real and diverse corner cases in marker-based auto-landing scenarios. Figure \ref{workflow} %\jz{still many issues with the figure. 1. Random Seed Generator shall be put insider the blackbox, and there shall be some Input Scenario Maps outside the box, you can take some from your figure 1.  2. The seed generator puts seed into the pool but from OnlineRL, once fitness function is used to assign seeds with fitness value you need to put seed back into the seed pool, you miss this link.  Also you don't need to use fitness function as a box, just use the box "Allocate seeds by fitness function values". 3. the arrow line from reinforcement leanring to next state is very ugly, position in the middle and don't overstrech the arrow end inside reinforcement learning box. 4. double check the font size, some are hard to read. If seed with.... if termination....and the three boxes inside surrogate environment training.}
%\jz{this figure is not what I want. I want you to create a workflow of GA from seed generation, seed selection, seed crossover and seed mutation. And put RL in the GA pipeline. It shall be a sequence diagram of how GA works, how GA works side by side with RL, why it is called RLaGA. You can even put Offline and online as two boundaries, so reviewers clearly know the intersection between offline and online} 
depicts the workflow of \tool. A seed generator initializes a set of random static environment seeds, each reflecting a specific test scenario. %It specifically considers the types of dynamic objects available that might affect UAV landing, as well as environmental conditions such as rain and weather. 
Each seed is input into the simulator, which then allows the RL agent to control NPC objects that interact with the UAV landing process online. Once the termination condition is met, each seed is assigned fitness values for each objective. These seeds are then varied, sorted and selected offline based on the NSGA-II multi-objective GA process to produce new static environment seeds \cite{deb2000fast}. The workflow terminates when the test budget is exhausted. %Subsequently, they undergo crossover and mutation to form the next generation, continuing until the simulation budget is exhausted.

\subsection{Modeling Test Scenario}
\label{ec}

% We aim to model the real-world marker-based landing scenario in the simulation. 
The following factors need to be modelled in the test scenario: UAV landing task, static environment condition, dynamic-object maneuvers. The termination condition is met once the UAV lands or experiences a collision/system crash. The typical workflow of marker-based landing task is defined as follows:
\begin{enumerate}
    \item{The UAV is instructed to move to a specified GPS coordinate, referred to as the marker position in this paper, representing the approximate location of the target landing marker.}
    \item{The landing system attempts to detect the target marker during the transit flight.}
    \item{Once the marker’s position is confirmed, the transit flight stops, and the landing begins.}
    \item{The landing system controls the landing descent such that the UAV performs a precision landing on the marker.}
\end{enumerate}

% \subsection{Scenario Representation}\llf{answer to "What is the rationale behind the selection of parameters?"}

\begin{figure}
    \centering
    \includegraphics[width=1\linewidth]{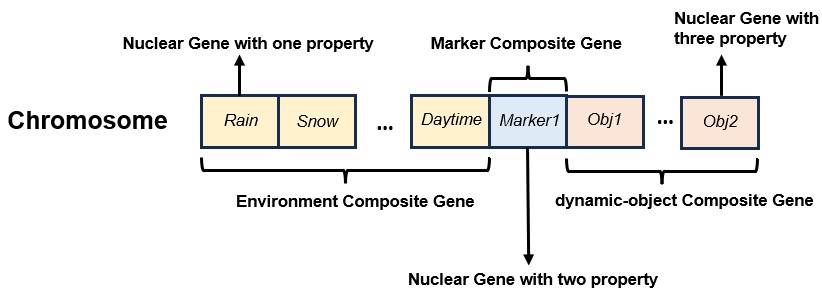}
	\caption{The Chromosome Representation of Test Case }
\label{genetic_representation}  
% \vspace{-0.8cm}
\end{figure}

% The static environment in our method can be represented by a chromosome which includes weather, daytime, marker position and available NPC-objects information of a test scenario. Figure \ref{genetic_representation} indicates the chromosome representation of a test scenario. The chromosome representation consists of an \textit{environment} composite gene, a \textit{marker} composite gene and a \textit{NPC-object} composite gene. The environment composite gene comprises single-scalar-property nuclear genes that denote the environment setting of specific environmental conditions in the test scenario, including \textit{dust, fog, rain, snow, road wetness, falling maple leaf, leaf on the road, snow on the road, and the position of the sun} in the simulation environment. The marker composite gene uses a two-scalar-property nuclear gene to represent the marker's (x, y) position. The NPC-object composite gene has several three-scalar-property nuclear genes to specify each NPC-object's \textit{type, start point and velocity}. A list of object types is specified by a given deployment scenario. For instance, for a deployment scenario, the available object types considered can be pedestrian, dog and bird. Moreover, NPC-object composite gene contains meta information to describe the count of each object type. The static environment is controlled by GA offline.
Figure \ref{genetic_representation} illustrates our approach, where a static environment is encoded as a chromosome, encapsulating elements like weather, daytime, marker position, and information about available dynamic objects within a test scenario. This chromosome structure includes three composite genes: \textit{environment}, \textit{marker}, and \textit{dynamic object}. The \textit{environment} composite gene comprises single-scalar nuclear genes representing specific environmental conditions, such as \textit{dust, fog, rain, snow, road wetness, falling maple leaves, leaves on the road, snow on the road}, and \textit{the sun's position}. These parameters can affect marker detection since we are testing vision-based landing systems. Dust, fog, rain, snow, and falling maple leaves can significantly impact the perception module relying on cameras. Snow and maple leaves on the road can hide the markers, while road wetness can slightly change the color of the ground, affecting detection results. The sun’s position affects the time of day and results in different lighting conditions. The \textit{marker} composite gene employs a two-scalar nuclear gene for the (x, y) coordinates of the marker's position. The marker position may cause corner cases, such as being under shadow or glare. The \textit{dynamic-object} composite gene contains multiple three-scalar nuclear genes to define each dynamic object's type, starting point, and velocity. The dynamic objects’ position and speed will result in different times to approach the UAV, affecting the online trajectory of the UAV, while the dynamic object's type may affect the detection and sensing result of the auto-landing system. Additionally, % each nuclear gene contains the number of this NPC-object as the meta information. 
the total number of nuclear genes in the \textit{dynamic-object} composite gene matches the variety of object types present in the given scenario. The GA manages this static environment configuration offline. The dynamic-object maneuvers indicate the behaviour of the dynamic object in the simulation. These are dictated by the online RL controls and include moving \textit{up, down, left, and right}.

%\jz{I would suggest you to change the dog, bird in Fig.3 to obj1, obj2, obj3 as the type information is stored as a property, and when you do crossover, you can easily calculate Type:count \lf{updated}.}
%, the number of nuclear genes is scenario-specified\jz{not scenario-specified. } .
%For example, at a specified test map, the involved dynamic objects can be pedestrian, dog and bird. 
 %The rationale for the parameter selection is based on the availability of UE4, we include all manipulable weather parameters, and essential marker and NPC-object information.

%For the property representing positions, the values are scaled to the range [0, 1] based on the available range of the test map.
%a nuclear gene in the dynamic object composite gene contains meta information to describe the count of the object.

%\jz{. In each deployment, the types of dynamic objects can be determined by a domain expert. I would suggest don't use dynamic object as the name for nuclear gene, you can give Dog, Cat, People, Car instead. And these nuclear gene has three properties, start, end, and the velocity. During your crossover and mutation, you can simply do a meta count to calculate for each type, how many available. Then based on the meta information, you can do crossover. Your mutation operation will be easier, as you just need to mutate each nuclear gene inside the dynamic object composite gene.}
%The start and end points, as well as the velocity for each dynamic object, are randomly sampled in every test case.
% \aftersubsectionskip

% \subsectionspacing
\subsection{Genetic Algorithm - Offline}
\label{GA}

In this paper, we present a new genetic algorithm for identifying diverse violation seeds for marker-based UAV landings. %We introduce a unique chromosome representation specifically for UAV landing scenarios, connecting offline GA and online RL.
Our design features a tailored multi-objective fitness function for our UAV auto-landing scenario, and specialized crossover and mutation operations to enhance seed diversity. %Further algorithm details are provided below. %\jz{I don't think it is still valid. As we just use GA to explore environment and the position of the marker.  We need to highlight our choromse representation, where we can also determine the initial number of dynamic objects and their types based on the deployment environment. Such Chromesome representation is specially designed and tailored for UAV landing scenario, this is our contribution. Another contribution is our dynamic object composite gene provides an interaction or connection for online RL. The online RL requires these dynamic object composite gene to intialize the state in the RL settings. This is our second contribution in terms of GA. I have also changed the subsection heading to reflect it is offline \lf{updated}}

\subsubsection{Multi-Objective Fitness Function}
\label{sec:multi_obj_fun}
Our multi-objective fitness function aims to identify varied scenarios that might compromise the safety of UAV landings. Compromises include landing in unmarked zones—areas that might hinder subsequent takeoffs or even lead to crashes. As such, the primary objective addresses incorrect landing positions. Additionally, extended landing durations are also risky due to the significant power consumption of UAVs. An increase in diversity among the population is also critical. Therefore, our multi-objectives can be identified as follows:

\begin{itemize}
    % \vspace{-0.6cm}
    \item \textit{Distance-To-Landing} (DTL), which measures the deviation from the designated landing marker to assess safety protocol compliance (objective 1).
    \item \textit{Time-To-Landing} (TTL), which evaluates the required landing duration to identify potential battery-related failures. For instance, the Wingcopter specification sheet indicates that their VTOL is limited to 2 minutes in multi-copter mode, necessitating that landing must occur within that time frame  \cite{Wingcopter_178_Heavy_Lift} (objective 2). 
    \item \textit{Diversity}, which reflects each seed's uniqueness, calculated by summing the distances between one seed and all other seeds in the generation to ensure a broad range of tested conditions, enhancing reliability. We adopt the novel search diversity calculation (Equation \ref{eq:parameter_distance-obj}) \cite{lehman2011abandoning,lehman2008exploiting}.
    %\vspace{-2cm}
    
    \begin{equation}
        D_i=\frac{1}{k} \sum_{j=1}^k d\left(x_i, x_j\right)
    \label{eq:parameter_distance-obj}
   % \vspace{-2cm}
    \end{equation}
where $D$ is the diversity metric for each chromosome representation, $x_i$ is the $i$-th chromosome representation of the test scenario in each generation, $x_j$ is the $j$-th chromosome representation of the test scenario in each generation, $d\left(x_i, x_j\right)$ is a function of calculating Euclidean distance between $x_i$ and $x_j$, and $k$ is the generation size (objective 3).

\end{itemize}

\par  
\subsubsection{Variation Operation} 
\label{uulc}

The \textit{crossover} and \textit{mutation} are used as variation operators. These operators, as indicated in Algorithm \ref{variation}, are used to manipulate chromosomes after one generation.

\begin{figure}
\centering
\includegraphics[width=1\linewidth]{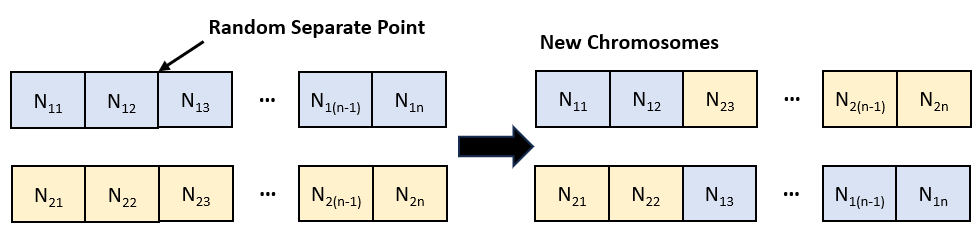}

\caption{Process of Nuclear Gene Crossover}
% \vspace{-3cm}
\label{crossoverdiagram}
% \vspace{-0.8cm}
\end{figure}

\begin{algorithm}
\begin{algorithmic}[1]
\caption{The GA chromosome-based suite of variation operators}
\label{variation}
    \STATE \textbf{Input}: Parents $P_t$, Offspring $O_t$, crossover threshold $threshold_c$, mutation threshold $threshold_m$, number of mutation candidates $m$
    \STATE \textbf{Output}: $P_{t+1}$, $O_{t+1}$
    \STATE{$P_{t+1}$ $\leftarrow$ $\emptyset$}, {$O_{t+1}$ $\leftarrow$ $\emptyset$}
    \STATE{$R_t$ $\leftarrow$ $P_t \cup O_t$}
    \FOR{i in range(0,$\lvert P_t \lvert$)}
        \STATE{sort and select parent chromosome $x_i \in R_t$}
        \STATE{$P_{t+1} \leftarrow P_{t+1} \cup \{x_i\}$}  
    \ENDFOR  
    \FOR{each pair of chromosomes $(x_i, x_j) \in P_{t+1}$}
    \STATE{generate $r \sim U(0,1)$}
        \IF{$r > threshold_c$}
        \label{cross1}
            \STATE{generate crossover point $s\sim U(0,Len(x_i))$}
            \STATE{$x_{i}^{'},x_{j}^{'}$ $\leftarrow$  $NuclearGeneCrossover(x_i, x_j, s)$}
            \STATE{$O_{t+1}\leftarrow O_{t+1} \cup \{x_{i}^{'}, x_{j}^{'}\}$}
        \ELSE
            \STATE{$O_{t+1} \leftarrow O_{t+1} \cup \{x_{i}, x_{j}\}$}
        \ENDIF
        \label{cross2}
    \ENDFOR
    \FOR{each chromosome $x_i \in O_{t+1}$}
        \FOR{each nuclear gene $y_i \in x_i$}
            \FOR{each property $y_{ij} \in y_i$}
                \STATE{generate $r \sim U(0,1)$}
                \IF{$r > threshold_m$}
                \label{mutationrate}
                    \STATE{$M \leftarrow \emptyset$}
                    \FOR{i in range(0,$m$)}
                    \label{candi1}
                        \STATE{generate $c \sim$ Property Range}
                        \STATE{$M \leftarrow M \cup \{c\}$}
                    \ENDFOR
                    \label{candi2}
                    \STATE{$y_{ij}^{'} \leftarrow PropertyMutation(y_{ij}, M)$}  
                    \STATE{$y_{ij} = y_{ij}^{'} $}
                    \label{mutationn}
                \ENDIF
            \ENDFOR
        \ENDFOR
    \ENDFOR
    \STATE{return $P_{t+1}$, $O_{t+1}$}
\end{algorithmic}
\end{algorithm}

\textbf{Nuclear Gene Crossover}: The crossover is a swap operation between two consecutive chromosomes. This aims to swap the nuclear genes between two high-fitness value chromosomes to form new chromosomes that potentially have high fitness values, as indicated in Figure \ref{crossoverdiagram}. If the crossover rate is met, the process begins by pairing two adjacent selected chromosomes. A random separation point within the chromosome length is chosen. The two paired chromosomes then disconnect from this point and crossover, forming two new chromosomes (line \ref{cross1} to line \ref{cross2} in Algorithm \ref{variation}).

% The first new chromosome retains a random sequence of consecutive nuclear genes from one original chromosome and acquires the remainder from its pair to become complete. The remaining nuclear genes from both in the pair constitute another new chromosome (line \ref{CROSSOVER1} to line \ref{CROSSOVER2} in Algorithm \ref{variation}).\jz{As discussed, please rewrite the paragraph to reflect we throw a dice from 1 to n-1, where n is the number of nuclear genes. The number will be used to separate each parent nuclear gene into two parts, head and tail. The head of the first parent is merged with the tail of the second parent while the head of the second parent is merged with the tail of the first parent. Give a quick figure to show this.  Highlight the number N is different for each deployment environment due to the types of dynamic objects available. You can change the name to Nuclear Crossover and make changes in the algorithm to reflect the name}

\textbf{Property-Level Mutation}: %The primary issue is that conventional genetic algorithms tend to converge on local minima and find similar solutions~\cite{lambora2019genetic, li2020av,schmidt2022stellauav}. Therefore, 
We design a mutation operation that would increase the diversity among chromosomes. Each property in the nuclear gene undergoes mutation according to the specified mutation rate (line \ref{mutationrate} in Algorithm \ref{variation}) which enhances exploration within the test scenario. Specifically, if the mutation rate is met, $m$ valid candidates for each property are randomly sampled based on the available range (line \ref{candi1} to \ref{candi2} in Algorithm \ref{variation}). Then, the candidate with the maximum distance to that property of all chromosomes in the current generation is set as the value of the property in the current mutated chromosome (line \ref{mutationn} in Algorithm \ref{variation}).

% \jz{As discussed, you can change the algorithm to reflect the name of mutation. The sequence is you will sample top K highest fitness value seeds, for each property for each nuclear gene, you will to find the mutation value which is as far away from the property values from the corresponding nuclear gene in all top K seeds. You can also quickly draw a figure to reflect that and remove the argmax formula which is quite confusing and hard to write it in a way easily understood by SE reviewers.}

\begin{equation}
\label{MAXID}
y_{ij}' = m_k , \quad k = \underset{k \in K}{\arg\max} \sum_{y \in Y_{ij}} |m_k - y|
\end{equation}

Equation \ref{MAXID} indicates our core mutation strategy, where  $y_{ij}^{'}$ is a $i$-th nuclear gene's $j$-th property in the chromosome after mutation, $K$ is the total number of mutation candidates, $Y_{ij}$ is a set containing all chromosomes' $i$-th nuclear gene's $j$-th property in the current generation. %\jz{I said before, writing formual requires very rigid form, k, you need to write k belongs to some set, here is the m, which is the total number of random sample. Also you apply L2 norm to the difference between two scalar which is also a scalar, why not just use absolute symbol. All the properties are scalars, right? not a vector}
% \aftersubsectionskip

% \subsectionspacing
\subsection{Reinforcement Learning - Online}

%In the context of marker-based UAV landing, nearby objects can unintentionally disrupt the UAV's landing, leading to violations. While GA-driven objects lack precise online control capabilities, introducing RL can amplify the chances of detecting such violations. 
 %the RL dynamically controls the maneuver of NPC-objects. 
Within the test scenario, RL guides the dynamic object dynamically in a direction that heightens the probability of violations using a reward function. This is done rather than sampling pre-defined trajectories for the dynamic objects as typically done in GA \cite{tian2022mosat}. %that reflects the percentage of the UAV camera's field of view that has been obscured.
Online RL testing faces challenges such as high training variance and long convergence times \cite{feng2023dense}. To mitigate these, we use a surrogate environment and a handcrafted reward function for preliminary RL training. The surrogate environment simplifies the learning process, reducing variance and speeding up convergence. Once trained, the agent is directly transferred to the full simulation.
\subsubsection{Build Surrogate RL Training Environment}

%Training RL agents in the full test process includes unnecessary and time-consuming simulations, such as the takeoff of the UAV and the animation of objects. To accelerate the training process, we built a more efficient surrogate training environment. 
The developed surrogate environment is built using the same simulation engine as the full simulation environment. Differently, the surrogate environment features a dynamic object, a marker, and a UAV. In the surrogate environment, the start point of the UAV is generated in the scenario and given the marker's coordinates as a prior. The UAV then flies directly to the marker without any additional trajectory planning. Each training episode begins with the UAV's takeoff and ends with either its landing or a collision. The following restrictions are applied to create a surrogate training environment with lower training variance:
\begin{itemize}
    % \item \textit{Camera mode} in the UE4 \cite{unrealengine} was used to simulate the flying and landing process of a UAV. The takeoff process occupies a substantial part of the simulation time. However, most violations occur due to in-flight marker detection and the landing process. The \textit{Camera mode} excludes the takeoff process, yielding significant time savings.
    \item In the surrogate environment, object animation effects, such as movement, are removed to expedite the process. The training focus is on enabling the RL agent to learn trajectory adjustment based on the UAV and marker positions, making animation effects unnecessary.
    % Objects move immediately to their respective states rather than displaying a smooth animation. During training, the goal is for the RL agent to learn trajectory adjustment based on the UAV and marker's position. Animation is not necessary in the training process.
    \item The marker is placed at a fixed point, greatly limiting the agent's exploration and thereby reducing training variance. Since the RL state is described using the relative distances between the marker, UAV, and object, the trained agent remains effective in the full environment.
    
    %Randomly spawning the marker and the object lead to a larger exploration space for RL, requiring more time to converge.
    \item 
    In the surrogate environment, a single dynamic object is deployed. After approximately 12 hours of training, the RL agent learns to produce trajectories towards the marker, impacting marker detection and drone landing. We then deploy multiple RL agents in the full environment to control several dynamic objects.
  %  \lf{In the surrogate environment, a single dynamic object is deployed. After about 12 hours of training, the RL agent learns to produce trajectories towards the marker, affecting marker detection and drone landing. We then deploy multiple RL agents in the full environment to control several dynamic objects.} %Once a single RL agent converges, multiple agents in the full simulation can use the same neural network weights to control various NPC-objects, as no communication is required among NPC-objects.

\end{itemize}
% After the single RL agent is trained in the surrogate environment, it is deployed in the main simulation environment where it can control multiple objects. 

\par The DQN algorithm is used %for the RL algorithm 
 as it is specifically designed for discrete action spaces \cite{van2016deep}. The state ($S$) of our RL input is a 4-dimensional vector that represents relative positional information between the object, the marker, and the UAV:

% \jz{it seems the relative positional information between the object and marker, marker and the UAV. You need to be clear in the surrogate simulation, you have only one dynamic object to control}:
\begin{equation}
\begin{split}
    S = (& P_{obj,x}-P_{marker,x}, P_{obj,y}-P_{marker,y}, \\
           & P_{uav,x}-P_{marker,x}, P_{uav,y}-P_{marker,y} )
\end{split}
\end{equation}
where $P_{obj}$ and $P_{uav}$ are the positions of the object and the UAV respectively, and $P_{marker}$ is the position of the marker. The definition of the state space is based on the rationale that relative positions more easily transfer between surrogate and full environments. 

The dynamic object's action space is discretized into categories of movement. The action space ($A$) can be denoted as a set that contains $5$ actions:
\begin{equation}
    A :=\{U,D,L,R,S \}
\end{equation}
where $U, D, L, R, S$ represents moving up, down, left, right, and stationary. The RL agent will choose one action from the whole action space at each time step. As RL makes step-wise decisions at short intervals (0.5 seconds), the accumulated movements form trajectories closely mimic realistic scenarios. Different velocities are set for various object types to closely replicate real-world conditions.

\subsubsection{Define Reward Function} 
% \llf{"The RL reward function only considers (partial) blocking of the drone's camera (3.2.2). I had imagined based on the discussion earlier in the paper that colliding with moving objects was an important class of violations independent of whether the UAV could "see" the landing marker (or the object)."}
In the surrogate environment, our dynamic object have two goals: first, to disrupt the UAV's detection of the marker; second, to attempt collision with the UAV. Consequently, a two-part reward function has been designed. Firstly, a semantic segmentation map is collected from the UAV's camera, detailing the marker and the dynamic object, and calculating the percentage of the obstructed area of the marker. Secondly, a collision indicator awards a reward upon collision. %Given the rarity of violations, we also established an intermediate reward to encourage the dynamic object to move closer to the marker. 
The reward function at each time step is defined as follows:
% Different from D2RL \cite{feng2023dense}, which treats violation as a rare-event, we use the handcrafted reward function to represent our violation. The violation can be defined as two

% Our objective is to enable the object under control to maximize the percentage of UAV camera vision that is blocked.
%its occupancy on the marker from the UAV's view.
% Therefore, we collect the semantic segmentation map from the UAV's camera which contains the semantic information for the marker and the object. Then, we calculate the percentage of the UAV camera's field of view being obscured.
%the object's occupancy on the marker. 

\begin{equation}
\label{reward_fun}
     % \mathbf{R} = \frac{S_{marker}}{S_d}
     \operatorname{\mathbf{R}}= \begin{cases}\frac{S_{gt}}{S_d} + \mathbf{I}, & \text { if } S_d !=0  \\ \mathbf{I}, & \text { if } S_d = S_{gt} \text{ or } S_d =0 \end{cases}
\end{equation}
where $S_{gt}$ is %the ground truth area of the marker
the area of the marker, a ground truth retrieved from the simulator and $S_d$ is the detected area of the marker through the camera at each time step. If $S_{gt} = S_d$, this signifies that the marker is not occupied and results in no reward. Similarly, if $S_d$ = 0, it indicates that no marker has been detected which also results in no reward. $\mathbf{I}$ is a collision indicator with a predefined reward value (1 used in our experiments). When a collision happens, the reward will be issued, subsequently concluding the episode. This unique design for surrogate training ensures rapid convergence of the RL agent and efficient transfer between the surrogate and the full environment.

% To allow the RL to further benefit the GA, the fitness function for RL augmented GA is updated as follows:
% \begin{equation}
% \label{fitrl}
% % Fitness Value = B_v + T_L + B_f + \frac{1}{\mathbf{R} }
% \operatorname{FitnessValue}= \begin{cases}DTL + TTL + \frac{1}{\sum\mathbf{R} }, & \text { if } \sum\mathbf{R} !=0  \\ DTL + TTL , & \text { if } \sum\mathbf{R} =0 \end{cases}
% \end{equation}
% Where $\sum\mathbf{R}$ is the accumulated reward value. A lower accumulated reward indicates a lower camera view obstruction throughout this simulation run, which would could result in UAV landing failure. %occur in unique boundary cases. 
% $\frac {1}{\sum\mathbf{R}}$ is used to ensure that cases with a very low accumulated reward result in a high fitness value. %The case of $\sum\mathbf{R} = 0$ removed from the fitness calculation as this case should not give any reward, as described above. \km{Could you please check that this still makes sense.}
% \aftersubsectionskip

% \aftersectionskip

% \sectionspacing
\section{Experiment}
\label{exp}
% \subsectionspacing
\subsection{Research Questions}

The following research questions (RQs) were assessed to evaluate the performance of \tool: 
\begin{itemize}
    % \item RQ1: Can a RL agent trained in a surrogate environment work in a final environment and benefit the GA?\jz{I suggest you don't mention our surrogate environment, this is just the way we used to expedite the RL training. What reviewers really care is whether our trained RL + GA works overall. I would suggest to directly compare our approach with baselines here using different metrics including violation found, how many rounds needed to find K bugs, the violation diversity metrics}
    % \item RQ1: How effective is $RLAGA$ in generating diverse landing violations across a variety of landing systems and scenarios?
    % \item RQ2: How effective is $RLaGA$ to expose landing violations compared to existing state-of-the-art techniques?
    % \item RQ3: How realistic are the violations found in the simulation?
    \item RQ1: How effectively does the \tool \ evaluate different marker-based landing systems?

    \item RQ2: How effective is \tool \ at exposing landing violations compared to existing state-of-the-art techniques?

    \item RQ3: Can landing violations found in the simulation be verified in the real world?
    
    % Can more diverse test cases be generated than the baseline methods in a given time budget? \jz{replace this one with ablation study, and analyse each compnent's importance here}
    % \item RQ2: How realistic are the violations found in the simulation? \jz{this is a good question, and we need to find those bugs only detectable by our approach and reproduced them in the real-world settings. In this RQ3,you need Alice to give detailed description of experiment evironment such as how to setup the test place and description of it and the uav used}
    % \item RQ3: How effectively does our approach test automatic landing systems? \jz{This is not a good question, you can say Can our method evaluate different UAV autoland systems? You need Sky to look at your description of your landing system. You don't need to put too much details, You can simply say a deployable OpenCV-based autoland system and another autoland system using YoloV5}
\end{itemize}
% \aftersubsectionskip

% \subsectionspacing
\subsection{Experiment Setup}

% \begin{figure}
% \centering
% \includegraphics[width=1\linewidth]{img/test_system.png}
% \caption{Testing System Architecture}
% \label{system}
% \end{figure}

% Figure~\ref{system} illustrates the overall architecture of our testing framework, consisting of three main components.  The simulation environment contains the environmental map or scenario to be simulated as well as the dynamic objects within that scenario. The auto-landing system incorporates modules for detection, planning, avoidance, and control. Our system utilizes Ardupilot \cite{ardupilot}, an industrial-grade UAV control platform, known for its high complexity. This, along with the modules' interoperability, contributes to the auto-landing system's complexity. The \tool\ evaluates the integrated system using both offline and online search methods.

% consists of the software to be tested, which can include machine vision marker detection, path planning, and obstacle avoidance.  The \tool \ algorithm controls the simulation parameters and searches for corner test cases that cause landing violations. 

\subsubsection{Simulation Environment}

In experiments, the simulation environment is constructed using AirSim~\cite{AirSim2017fsr}, which is an open-source UAV simulation platform based on the Unreal Engine (UE) 4~\cite{unrealengine} and widely used in UAV research~\cite {lai2022ai, song2021autonomous, shimada2022pix2pix}. AirSim supports rendering high-fidelity scenes and simulating realistic physical and weather effects. Another advantage is that it is flexible to customize and create new simulation maps leveraging resources in UE4. We can also easily extend the ability of AirSim by implementing new APIs to manipulate the simulation environment. 

\begin{figure}
\centering

\begin{subfigure}[b]{0.23\textwidth}
\centering
        \includegraphics[width=.8\textwidth]{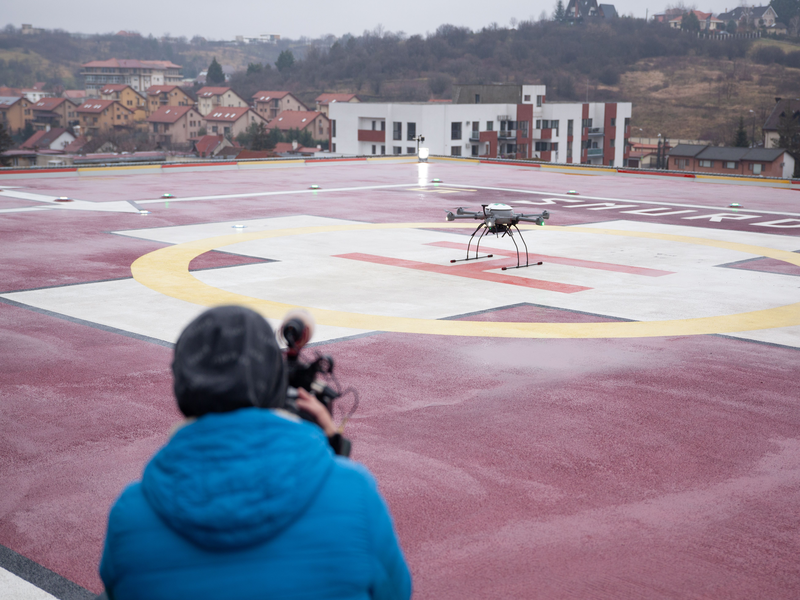}
        \includegraphics[width=.8\textwidth]{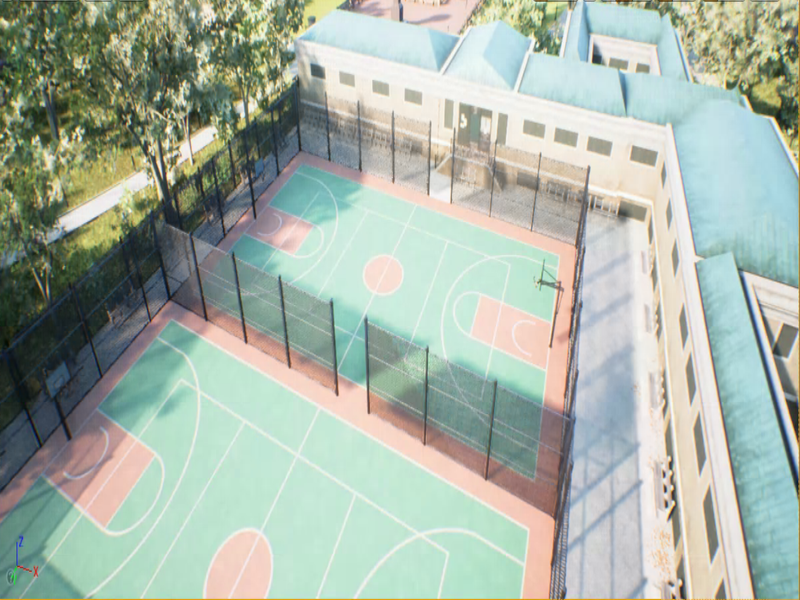}
        \caption{Court}
        \label{Ground}
\end{subfigure}
% \begin{subfigure}[b]{0.23\textwidth}
%         \includegraphics[width=\textwidth]{img/ske_map/lake_reals.png}
%         \includegraphics[width=\textwidth]{img/ske_map/lake_sims.png}
%         \caption{Lake}
%         \label{Lake}
% \end{subfigure}
\begin{subfigure}[b]{0.23\textwidth}
\centering
        \includegraphics[width=.8\textwidth]{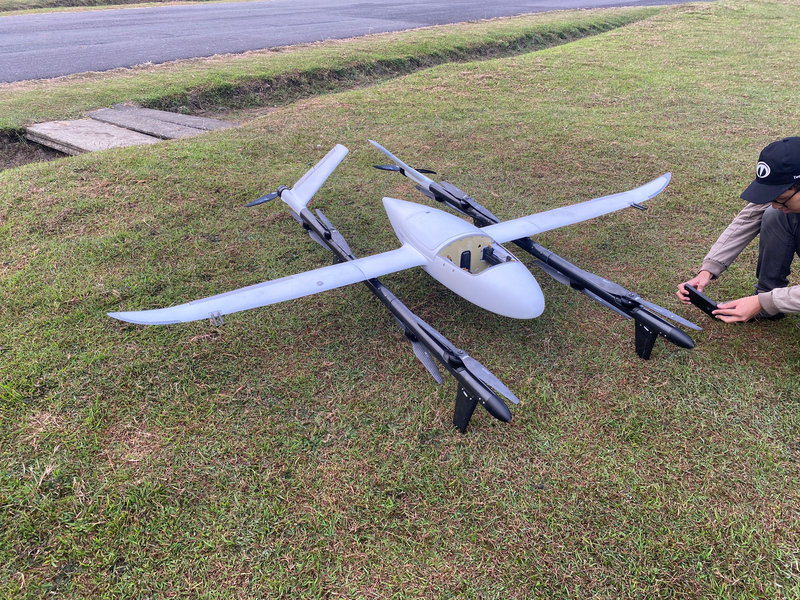}
        \includegraphics[width=.8\textwidth]{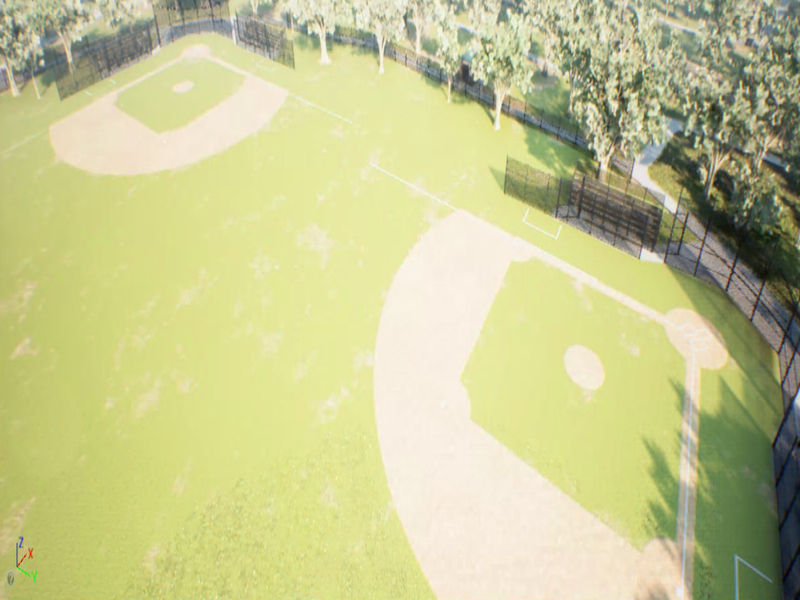}
        \caption{Lawn}
        \label{Lawn}
\end{subfigure}
% \begin{subfigure}[b]{0.2\textwidth}
%         \includegraphics[width=\textwidth]{img/ske_map/garden_reals.png}
%         \includegraphics[width=\textwidth]{img/ske_map/garden_sims.png}
%         \caption{Garden}
%         \label{Garden}
% \end{subfigure}
\caption{Example of test map in real-world (First row) and AirSim (Second row).}

\label{testmap}
% \vspace{-0.4cm}
\end{figure}

Two maps based on the real-world test fields were chosen by industry research collaborators to construct realistic simulation test environments, as shown in Figure~\ref{testmap}. The \texttt{Court} map features a city-like environment with basketball fields, %man-made structures, 
including concrete landing areas, while the \texttt{Lawn} map offers a natural setting. In addition, ArUco markers, a widely deployed type of UAV-landing target, are inserted as UAV landing targets (``markerID:0''), with an API is provided to place the marker \cite{garrido2014automatic, khazetdinov2021embedded}. The selected dynamic object types aim to replicate the variety of dynamics typical in those scenarios, such as aerial movements, low-height ground activities, and interactions involving taller ground elements. Therefore, different types of dynamic objects were added on the ground and in the air of the simulation maps, including persons, birds, and dogs.  Corresponding APIs were also implemented to control the dynamic movements. The simulation environments and new APIs can be downloaded from the available software repository (see Abstract).

%For different simulation maps, we set different types of dynamic objects based on the characteristics of the maps. For the map \texttt{Ground}, only \textit{persons} is set in the map. For the map \texttt{Neighborhood}, \textit{persons}, \textit{birds}, and \textit{dogs} are set. For the map \texttt{Lawn}, the three dynamic objects and \textit{zebras} are set in the simulation environment.   

\subsubsection{Marker-based Landing Systems}

 % The auto-landing system we cuincorporates modules for detection, planning, avoidance, and control. Our system utilizes Ardupilot \cite{ardupilot}, an industrial-grade UAV control platform, known for its high complexity. This, along with the modules' interoperability, contributes to the auto-landing system's complexity. 
 
\begin{figure}
\centering

% \begin{subfigure}[b]{0.45\textwidth}
% \centering
\includegraphics[width=0.45\textwidth]{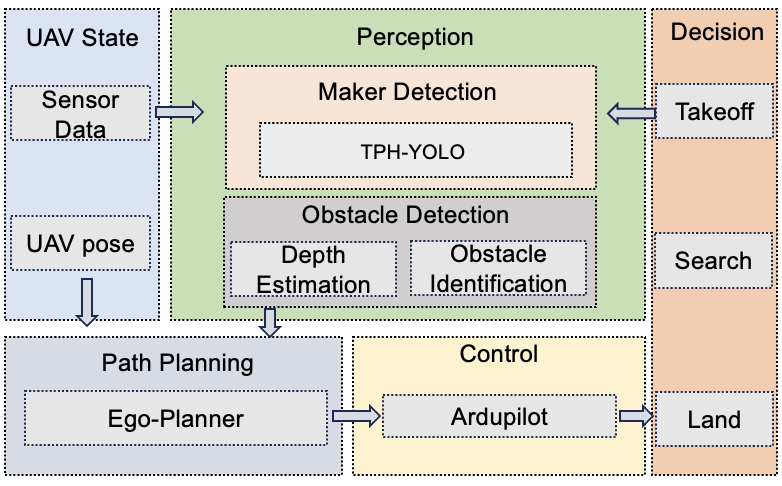}

% \end{subfigure}
% \begin{subfigure}[b]{0.23\textwidth}
%         \includegraphics[width=\textwidth]{img/ske_map/lake_reals.png}
%         \includegraphics[width=\textwidth]{img/ske_map/lake_sims.png}
%         \caption{Lake}
%         \label{Lake}
% \end{subfigure}
% \begin{subfigure}[b]{0.45\textwidth}
% \centering
%         \includegraphics[width=0.8\textwidth]{img/tphopencv.png}
      
%         \caption{TPHYoLo-MLS and OpenCV-MLS}
%         \label{tphopencv}
% \end{subfigure}
% \begin{subfigure}[b]{0.2\textwidth}
%         \includegraphics[width=\textwidth]{img/ske_map/garden_reals.png}
%         \includegraphics[width=\textwidth]{img/ske_map/garden_sims.png}
%         \caption{Garden}
%         \label{Garden}
% \end{subfigure}
\caption{ The system architecture of MM-MLS}
%\vspace{-0.5cm}
\label{system}
% \vspace{-0.5cm}
\end{figure}

\begin{table*}[h!]
\centering
\caption{The component involved in each SUT}

\begin{tabular}{|c|c|c|c|c|c|c|c|}
\hline
\textbf{Landing Systems} & \textbf{UAV State} & \multicolumn{2}{c|}{\textbf{Detection}} & \textbf{Planning} & \multicolumn{2}{c|}{\textbf{Control}} & \textbf{Decision} \\
\cline{3-7}
 &  & OpenCV \cite{opencv_library} & TPHYolo\cite{zhu2021tph} & Ego-Planner\cite{zhou2020ego} & Airsim \cite{AirSim2017fsr} & Ardupilot\cite{ardupilot} &  \\
\hline
OpenCV-MLS  & \Checkmark & \Checkmark &  &  & \Checkmark &  & \Checkmark \\
\hline
TPHYolo-MLS & \Checkmark &  & \Checkmark &  & \Checkmark &  & \Checkmark \\
\hline
MM-MLS & \Checkmark &  & \Checkmark & \Checkmark &  & \Checkmark & \Checkmark \\
\hline
\end{tabular}

\label{system_comp}
% \vspace{-0.5cm}
\end{table*}
Three marker-based landing systems were implemented and integrated into the simulated UAV and tested in the created simulation environments (Table \ref{system_comp}): 
\begin{itemize}
    \item \textit{OpenCV-MLS}: The system features a perception module utilizing OpenCV \cite{opencv_library} for marker detection and position estimation, a decision module to manage the marker-based landing task, and an Airsim-based control module.  %Then, the drone is controlled to fly toward the marker along a straight-line trajectory.}
    \item \textit{TPHYolo-MLS}: The architecture mirrors \textit{OpenCV-MLS} but incorporates TPHYolo~\cite{zhu2021tph}, a deep learning model for marker detection. TPHYolo adopts the transformer mechanism~\cite{vaswani2017attention} to improve performance when detecting small-size objects, as proven for use in UAV-related research~\cite{zhu2021tph}.
    % an implementation of the TPHYoLo~\cite{zhu2021tph}  deep learning model as the marker detection algorithm based on \textit{OpenCV-MLS}. 
    \item \textit{MM-MLS}: This system is built based on \textit{TPHYolo-MLS}, and introduces ego-planner~\cite{zhou2020ego} for trajectory planning. Ego-planner, a leading open-source solution, enables real-time obstacle detection and trajectory planning using depth image data and UAV's pose. The ArduPilot \cite{ardupilot} flight control platform directs the UAV to follow the planned trajectory. The detailed system architecture, illustrated in Figure \ref{system}.%The architecture of this system is indicated in Figure \ref{TPHMM}.

    % a multi-module landing approach, employing TPHYoLo for marker position detection and fast-planner~\cite{zhou2020ego} for trajectory planning.  (see Figure \ref{TPHMM}).
    % This is the most fully featured landing system as it includes obstacle avoidance.}
\end{itemize}

% Each of the planners were chosen for their ability to achieve real-time obstacle detection and trajectory planning on commercially-available computing hardware that is suitable for use on small drones. 

Though some other marker-based landing systems have been proposed~\cite{lin2021real, brunner2019urban}, their models were not available as open-source at the time of writing and, thus, were not evaluated in this study.

\subsubsection{Implementation Details}
% \begin{table}[!ht]
% \centering
% \caption{RL Agent, GA and Simulation Hyperparameters}
% \label{tab:hyperparameters}
% \resizebox{\columnwidth}{!}{
% \begin{tabular}{lll}
% \hline
% \textbf{Component} & \textbf{Parameter} & \textbf{Value} \\
% \hline
% RL Agent (DQN) & Neural network layers & 2 \\
% & Neurons per layer & 24 \\
% & Learning rate & 0.001 \\
% & Initial epsilon & 0.99 \\
% & Epsilon decay & 0.995 \\
% & Minimum epsilon & 0.01 \\
% & Collision reward ($\mathbf{I}$) & 1 \\
% & Training episodes & 800 \\
% % & Training duration & 4 hours on RTX3090 GPU \\
% \hline
% GA & Population size & 20 \\
% & Generations per trial & 20 \\
% & Simulation Budget & 400 \\
% & Mutation rate & 0.2 \\
% & Crossover rate & 0.2 \\
% & Mutation candidates & 10 \\

% \hline
% Simulation Setting & Positional parameter range & [-1,1] \\
% % & Weather parameter default range & [0,1] \\
% & Weather parameter confined range & [0,0.15] \\
% % & NPC-object number & By PC configuration \\
% % & NPC-object movement distance & 2 units \\
% \hline
% \multicolumn{3}{l}{\textit{}} \\
% \end{tabular}}
% \end{table}

%Table \ref{tab:hyperparameters} indicates 
The detailed hyper-parameter setting of our method and the experiment can be found in our supplementary material. Specifically, in simulations, positional parameters are normalized within [-1,1], while weather parameters default to the [0,1] range. To ensure moderate weather conditions, these parameters are confined to [0,0.15]. 

The number of dynamic objects in the simulation varies with the PC's capabilities. To control for variability due to simulation rendering differences, all experiments are repeated three times using an RTX 3090 GPU, each with a 400-test budget, and average results are reported. The UAV's dimensions are set to one meter in width and length, matching AirSim's default configuration. The target marker is 1.5 meters in width and length, allowing the UAV to land within it if the auto-landing system functions correctly.

\subsection{Experiment Design}
\label{design}

\subsubsection{RQ1} %The ability of the \tool \ to generate diverse violations is evaluated. 
We assessed \tool \ on three landing systems: \textit{OpenCV-MLS}, \textit{TPHYolo-MLS}, and \textit{MM-MLS} in two simulation maps \texttt{Court} and \texttt{Lawn} respectively.
Each landing system was assessed by using \tool \ to generate 400 test scenarios. Data for each scenario run was saved, including the scenario seed, a video recording, the UAV trajectory data, and any UAV-object collision events.

% {\color{blue} Real-world performance of a UAV with only GPS input has a position estimate that is typically accurate within than 1 meter of the truth position \cite{fornasierequivariant}. Addition visual odometry estimation data, such as a that provided by the autoland software, allow the UAV's estimated position to be within 0.05 meters \cite{kalaitzakis2021fiducial} and controlled position to be within a similar range of error \cite{Brescianini2013NonlinearQA}}. 

The record was used to identify landing violations of the tested system. A violation occurred if the UAV collided with objects or landed outside the marker's bounds. Given the marker is a 1.5-meter square and the UAV is smaller, a landing position more than 1.5 meters from the marker center ensures it is outside the marker. Considering real-world UAV GPS accuracy is typically within 1-meter \cite{fornasierequivariant}, we set 1.5 meters as the threshold for landing violations. The \textit{landing violation percentage} was calculated as the ratio of scenarios with violations to the total number of scenarios. This value measures the effectiveness of the \tool \ method in finding landing violation scenarios.
%The record was used to identify whether a test scenario caused a landing violation of the tested landing system. A landing violation was said to occur either when the UAV collides with other objects, or when the landing is finished and the UAV's position is outside the bounds of the landing marker. As the size of the marker is a 1.5-meters square and the UAV size is smaller than the marker size, the distance between the landing position and the marker center greater than 1.5 meters can guarantee the UAV is outside of the marker. Meanwhile, real-world performance of a UAV with only GPS input has a position estimate that is typically accurate within 1 meter of the truth position \cite{fornasierequivariant}. Therefore, we set 1.5 meters as the threshold to check for violation landings. 
% A marker size of 1.5 meters square was selected to meet real-world performance expectations. A violation occurred when the distance between the drone center and the marker center is larger than 0.75 meters. This tolerance means that GPS-only landings are likely to cause a violation where as visually assisted landings should almost never cause a violation if operating correctly. 
%A \textit{landing violation percentage} was calculated as the ratio of scenarios with violations divided by the total number of scenarios. This value measures how effective the \tool \ method is at finding landing violation scenarios.

The saved landing recordings were also analyzed to identify the \textit{violation type}, which qualitatively measures the diversity of violation scenarios that are caused by different reasons.
% generated test scenarios. The reason for the landing failure is attributed to a specific type of violation. %The \textit{violation type} is the reason for the failed landing. 
For example, the reason could be false positive marker detection, collision, or even system crashes. The metric \textit{violation type} showcases the \tool's capability to create a diverse range of landing violation scenarios.

%A large number of unique violation causes demonstrates the ability of the \tool \ method to generate diverse landing violation scenarios. 

\subsubsection{RQ2} We compared the performance of  \tool \ with five baseline methods on \textit{MM-MLS}, the system currently deployed in the real-world by our industry collaborators, in two simulation maps. The detailed settings of those baselines are listed below:
\begin{enumerate}

    \item \textit{Random}: all the static environments are randomly generated. Additionally, dynamic objects' start positions and destinations are generated randomly within the specified map range.
    
    \item \textit{Multi-Objective GA}  \cite{tian2022mosat}: This baseline shares the same fitness function and chromosome representation as \tool, employing NSGA-II \cite{deb2000fast} to manage these objectives. However, this baseline adds the dynamic object's destination into the dynamic object composite gene and utilizes a standard mutation strategy. This baseline can also be regarded as an ablation study to indicate the performance of GARL without RL part.

     \item \textit{Offline RL Fuzzer}  \cite{bottinger2018deep}: 
     this baseline initiates with a random scenario, and the state encodes the environment, marker position, and dynamic objects' start and endpoints. The RL agent can alter weather conditions or marker location by ±0.1 or move a dynamic object's start or endpoint in one of four directions: up, down, left, or right offline. The reward combines Distance to Landing (DTL), Time to Land (TTL), and diversity to all previously generated scenarios as introduced in Section~\ref{sec:multi_obj_fun}. The configuration for the offline RL model, including its exploration strategy (initial epsilon and its decay) and learning rate, parallels our surrogate training protocol. Detailed settings of this baseline can be found in our supplementary material.
    \item \textit{Online RL}  \cite{lu2022learning}: this baseline allows the RL agent to control weather, daytime, and dynamic objects online. Each simulation starts with a randomly initialized test scenario. The state encodes the environment, marker position, and dynamic objects' start points. The RL agent can alter the weather by ±0.01 or move a dynamic object up, down, left, or right. The reward function mirrors that of our surrogate environment. If the action involves moving a dynamic object, a random object will be selected. Detailed settings are provided in our supplementary material.

   % this baseline allows the RL agent to control both weather, daytime, and dynamic objects online. Each simulation is started with a randomly initialized test scenario. The state encodes the environment, the marker position, and available dynamic objects' start points. In the simulation, the RL agent can alter a weather condition by ±0.01 or direct an dynamic object in one of four directions: up, down, left, or right online. The reward function mirrors that of our surrogate environment.  If the action is to move the dynamic object, a random dynamic object will be selected to move. The detailed settings of this baseline can be found in our supplementary material.

    % \begin{equation}
    % \begin{split}
    %     S_{online} = ( & dust, fog, \ldots, snow, \\
    %                    & P_{obj,x}-P_{marker,x}, P_{obj,y}-P_{marker,y}, \\
    %        & P_{uav,x}-P_{marker,x}, P_{uav,y}-P_{marker,y}) 
    % \end{split}
    % \label{online_state}
    % \end{equation}
    % \begin{equation}
    % \begin{split}
    %     A_{online} := \{ & dust\pm0.01, \ldots, snow\pm0.01, \\
    %     & U, D, L, R, S \}
    % \end{split}
    % \label{online_action}
    % \end{equation}
    
    % where $P_{obj}$, $P_{uav}$, and $P_{marker}$ indicate the position of NPC-object, UAV and marker.

    \item \textit{Surrogate trained RL with random}: this baseline serves as an ablation study, replacing the GA offline component in our solution with a random generator.
    
    %this baseline can be regarded as the ablation study, which is replaced the GA offline component in our solution with a random generator.  
    
    %This baseline can be regarded as the ablation study, which is built based on the Random baseline. Differently, this baseline allows the RL agent trained from the surrogate environment to control the NPC-object in the scenario.

\end{enumerate}
% The first baseline method is
% \textit{Random}, where all scenario seeds in all generations are randomly sampled, and the second baseline method is \textit{Many-Objective GA}, where the search objectives are the same with \tool, and non-domanient sorting is also applied on manage those objectives \cite{panichella2015reformulating, abdessalem2018testing, tian2022mosat}. Instead, the In addition, to evaluate the effectiveness of the RL component in the proposed method, we conducted an ablation study by only using the GA (\textit{Diverse GA}) component in the proposed method as another baseline to compare.
% An ablation study was also applied to evaluate the  {\color{blue}introduce the ablation study setting}\km{Don't forget this}.

%We compared the efficiency of all methods to generate landing violation scenarios and quantitatively measured the diversity of generated test scenarios for all methods.
The efficiency metric is measured as \textbf{Top-K}~\cite{deng2022scenario}, which evaluates how many test cases are needed to find the first \textit{K} violation cases. %test scenarios cases that cause landing violations are generated. 
In this paper, the K is set as 10. The \textit{landing violation percentage} metric, used in RQ1, was used to evaluate the percentage of violation in generated test cases. %how many of the generated scenarios detect landing violations. %in this unit of time. 
Diversity was measured with two metrics: \textit{parameter distance} and \textit{3-D trajectory coverage}.

\textbf{Parameter distance} was derived from the idea of novel search~\cite{lehman2011abandoning,lehman2008exploiting}.  A higher parameter distance means the generated violation seeds have a higher diversity. Parameter distance is calculated for all generated test cases as shown in Equation~\ref{eq:parameter_distance}:
\begin{equation}
    \rho(x)= \frac{1}{n} \sum_{i=1}^n\left[ \frac{1}{n} \sum_{j=1}^n d\left(x_i, x_j\right) \right]
    \label{eq:parameter_distance}
\end{equation}
where $x_i$ and $x_j$ are the $i$-th and $j$-th test scenario representation vector in all generated scenarios, which contains the environment (weather, daytime), the marker position, and available dynamic object information,  $d(x_i,x_j)$ represents the Euclidean distance between $x_i$ and $x_j$, $n$ is the number of generated test cases.

\textbf{3D-Trajectory Coverage} is a trajectory coverage metric that measures how violated trajectories cover the simulation map, and is adapted from an ADS implementation \cite{hu2021coverage}. 
We implement the metric in 3D in order to measure the trajectory coverage during a 3D marker-based UAV landing task. The simulation environment is divided into several $2\times2\times2$ meter cells. A 3-dimensional array is used to index each cell in the grid. All entries in the array are initialized to \textit{false}. An entry's value is set to \textit{true} if the UAV visits the corresponding grid cell while a violation occurs during that run. The metric calculation is presented as Equation \ref{coverageeq}:
\begin{equation}
    3D-Trajectory Coverage = \frac{N_{covered}}{N_{total}}
    \label{coverageeq}
\end{equation}
where $N_{covered}$ represents the number of grids covered in violation cases, $N_{total}$ denotes total number of cell.

\par Additionally, we measure the time consumption of each method. While accuracy is crucial, significantly longer test case generation can hinder successful adoption.
\subsubsection{RQ3} 
%All identified violation types that can be safely reproduced were replicated in real-world scenarios based on simulation records. 
Real-world testing was conducted to ensure the generated violation cases were realistic. The \textit{MM-MLS} landing system was implemented on a custom UAV to recreate the simulated errors. This UAV is equipped with a 295mm frame, 5-inch 3-blade propellers, 1750KV motors, an Intel D455 camera, and an NVIDIA Jetson Nano onboard computer. A physical ArUco marker (``markerID:3") serves as the landing target on the ground. For non-collision violation types, the UAV is programmed to initiate landing after detecting the marker continuously for 5 seconds above a 0.7 confidence threshold. For reproducing collision-related violations, we utilize a controlled real-world replication strategy. Specifically, we deployed the auto-landing system on the UAV. The system would output the planned trajectory for landing, but the UAV keeps hovering in the air and does not follow the trajectory to move. The intermediate outputs of the system such as detected obstacles and planned trajectories are recorded and can be visualized in a visualization software called Rviz~\cite{kam2015rviz} to check the reaction of the landing system regarding obstacles.
% This involves preventing the drone from carrying out actual landing maneuvers. Instead, we visualize the planner's trajectory to show the how landing system will react to the obstacle. 
By comparing the planned trajectory, we can predict if collisions in simulations would occur under real-world conditions, allowing for a safe assessment of potential collision scenarios without risking actual damage.
\section{Results}
\label{result1}

% \subsectionspacing
\subsection{Effectiveness of \tool \ across different Auto-Landing systems (RQ1)}
% \subsubsection{Effectiveness of generating landing violation scenarios}
% {\color{blue} Write result analysis for Table~\ref{Ground_system}}
% \begin{table*}
%     \centering
%     \scalebox{0.9}{
%   \begin{tabular}{c|c|c|c}
% \hline & $OpenCV$ & $TPHYoloV5$ & $Multi-Module$ \\
% \hline Landing violation types & 6 & 11 \\
% \hline $\begin{array}{c}\text { Number of test cases generated } \\
% \text {to find one violation}\end{array}$ & 93 & 28 \\ 
% \hline $\begin{array}{c}\text { Number of test cases generated } \\
% \text { to find all violation types }\end{array}$ & $15 \mathrm{~h}$ & $33 \mathrm{~h}$ \\
% \hline
% \end{tabular}
% }

% \caption{Comparison results across different landing systems on the Ground map. {\color{blue}In the table, the result for Multi-module MLS may be the average result on three maps}}
% \label{Ground_system}
% \end{table*}

\begin{table}[]
\centering
\caption{Landing violation percentage for different landing systems across two maps}
\begin{tabular}{rccc}
\hline
\multicolumn{1}{l}{\textbf{}} & \multicolumn{1}{l}{\textit{OpenCV-MLS}} & \multicolumn{1}{l}{\textit{TPHYolo-MLS}} & \multicolumn{1}{l}{\textit{MM-MLS}} \\ \hline
Map Court                    &               71.50\%                       &                                        30.96\%  &     20.60\%                              \\ \hline
Map Lawn                      & 42.75\%                                       &38.25\%                                       & 17.11\%                                   \\ \hline

\end{tabular}

\label{tab:violation_percentage}
\end{table}

\begin{figure}
\centering
\includegraphics[width=1\linewidth]{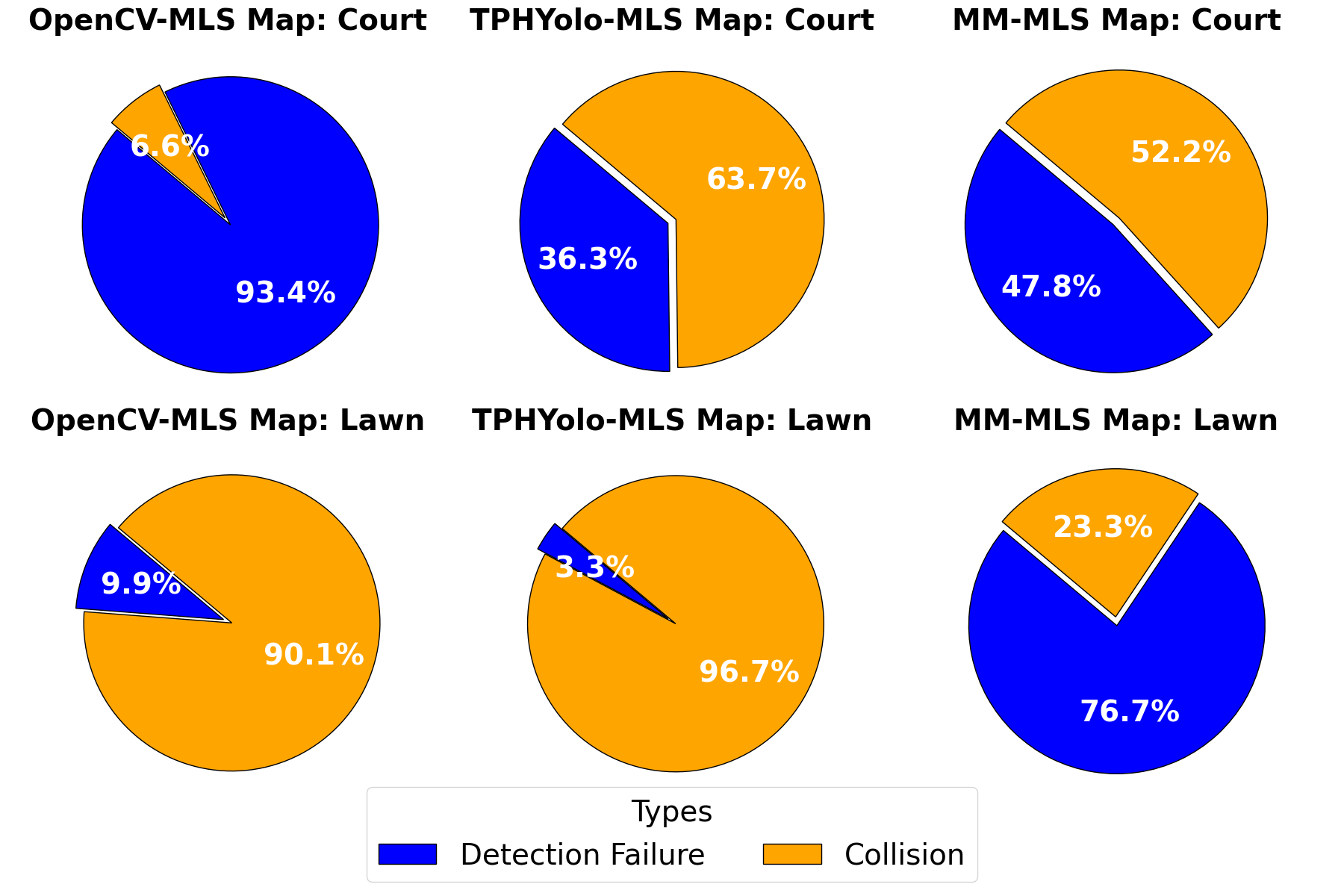}

\caption{General violation types found in different systems.}

% \vspace{-0.8cm}
\label{pte}
\end{figure}

%Table \ref{tab:violation_percentage} displays our method's test results across diverse landing systems and test maps, underscoring its efficacy in producing landing violations for examined industry evolving UAV landing systems. 
Table \ref{tab:violation_percentage} showcases the test results of \tool \ across three landing systems and different test maps. The results highlight \tool's effectiveness in identifying landing violations for these emerging versions of industry-standard UAV landing systems. More advanced versions of landing systems have led to a significant decrease in violations detected, with a notable %56.69\% 
40.54\%
decrease when moving from \textit{OpenCV-MLS} to \textit{TPHYolo-MLS} in the \texttt{Court} map. %highlighting the superior robustness of deep learning for marker detection. 
Furthermore, the transition from \textit{TPHYolo-MLS} to \textit{MM-MLS} saw a further %55.26\%
21.14\%
decrease in violations detected in the \texttt{Lawn} map.

% , underscoring the superior performance of this system. 

To compare system performances, we categorized violations into \textit{collision} and \textit{detection failure (of marker)}, shown in Figure \ref{pte}. The \texttt{Lawn} map, with its uniform green surface, mainly shows collision violations, with minimal interference in marker detection. In contrast, the \texttt{Court} map, featuring white lines, red areas, and tree roots, presents a greater chance of false positives, thus increasing marker detection failures.
%To explore performance differences among systems, we sorted the identified violations into two general categories: \textit{collision} and \textit{detection failure (of marker)}, as illustrated in Figure \ref{pte}. The \texttt{Lawn} map, characterized by its uniform green ground, predominantly exhibits collision violations due to minimal marker detection interference. This contrasts with the \texttt{Court} map, where additional elements like white lines, red area and tree roots increase the likelihood of false positives, leading to a higher rate of marker detection failure. %This also points out our future direction on training a more robust detection model. 
Therefore, in the \texttt{Lawn} map, \textit{OpenCV-MLS} and \textit{TPHYolo-MLS} perform similarly without a planning module. However, on the \texttt{Court} map, \textit{OpenCV-MLS} predominantly suffers from \textit{detection failure} due to its less robust detection algorithm, which limits the effect of dynamic objects controlled by \tool. Conversely, \textit{TPHYolo-MLS} experiences fewer \textit{detection failure} violations, thanks to its superior detection capabilities in detecting smaller objects, enabling more accurate marker identification. Yet, the absence of an obstacle avoidance module under the testing of \tool \ significantly raises the risk of dynamic object collisions with the UAV.

\begin{figure}[]
\centering

\begin{subfigure}[b]{0.5\textwidth}
\centering
\includegraphics[width=0.24\linewidth]{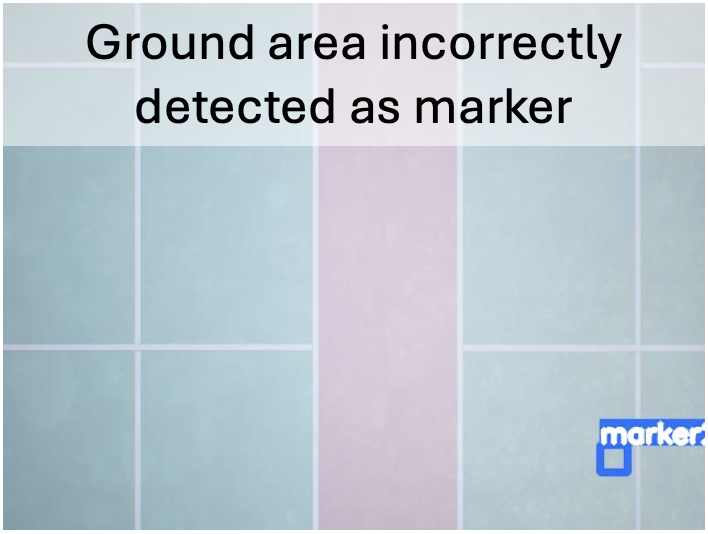}
\includegraphics[width=0.24\linewidth]{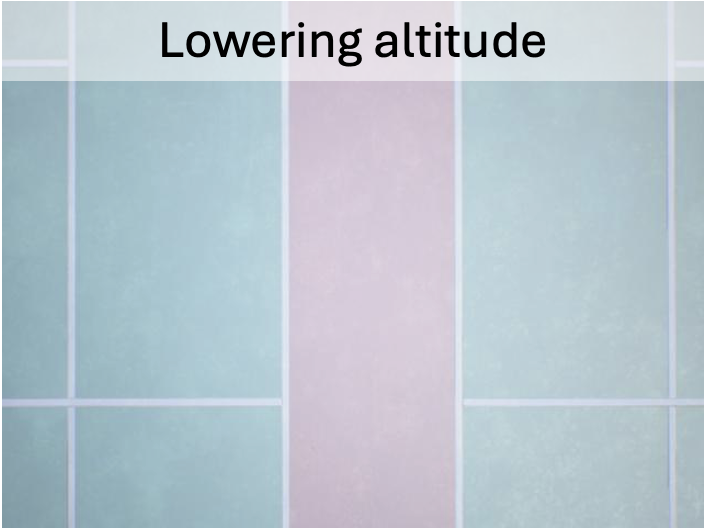}
\includegraphics[width=0.24\linewidth]{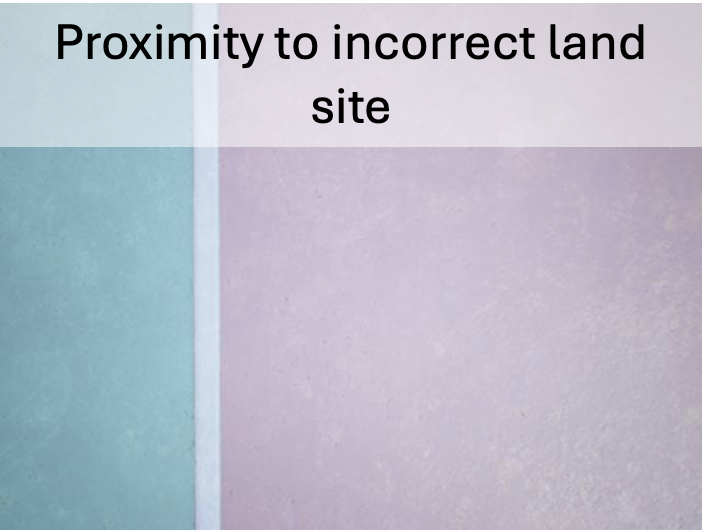}
\includegraphics[width=0.24\linewidth]{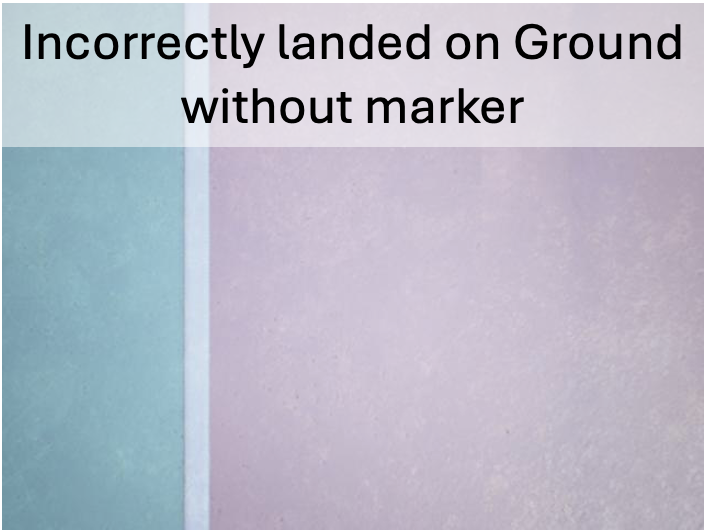}
\includegraphics[width=0.24\linewidth]{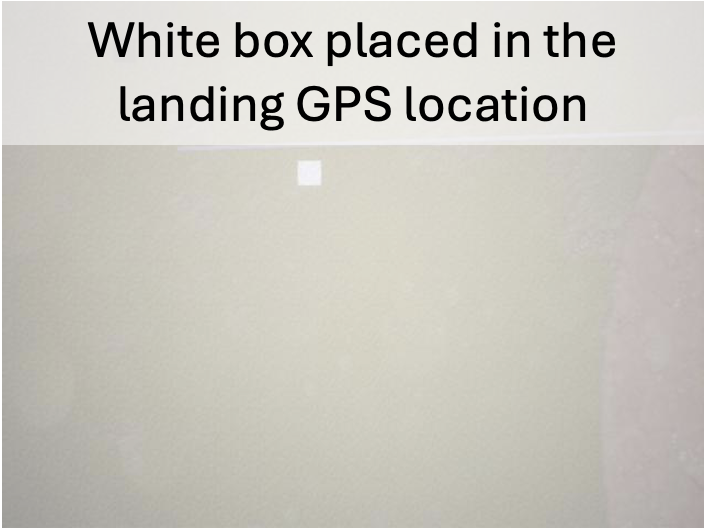}
\includegraphics[width=0.24\linewidth]{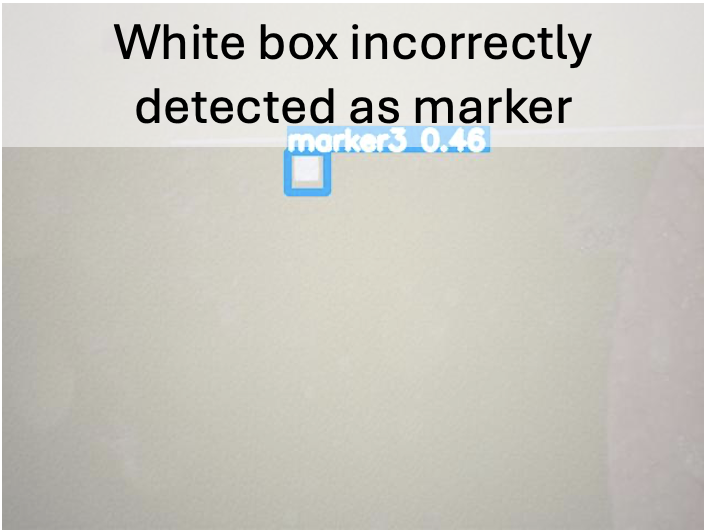}
\includegraphics[width=0.24\linewidth]{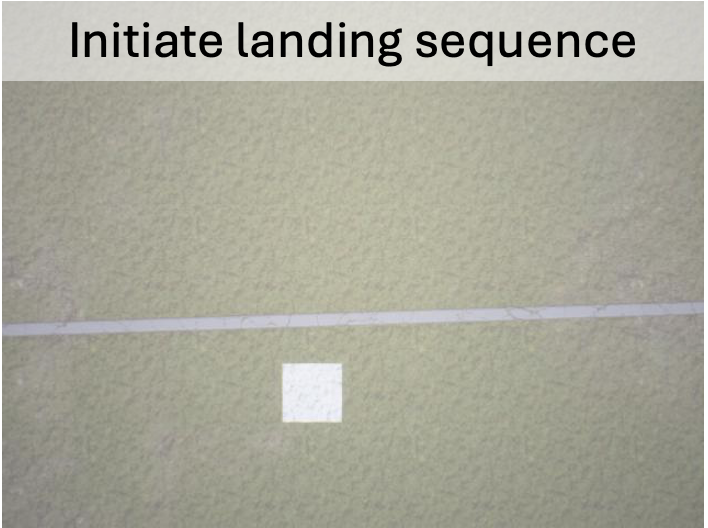}
\includegraphics[width=0.24\linewidth]{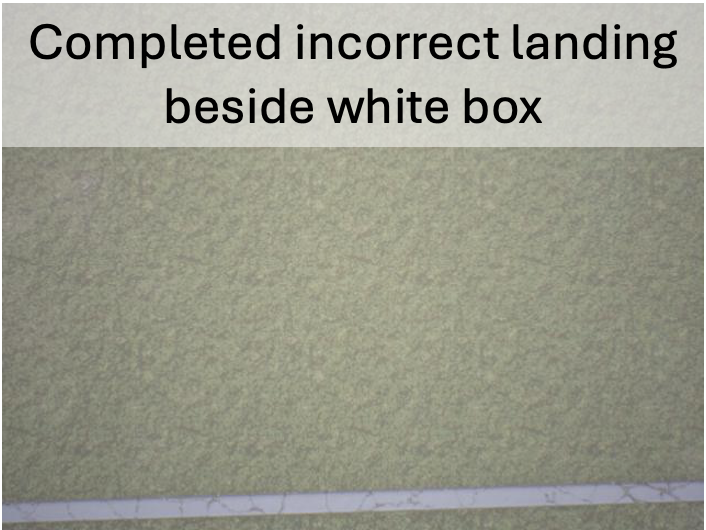}
\caption{Examples of Violation Type I: UAV lands on a false positive location due to wrong detection}
\label{example1}
\end{subfigure}

\begin{subfigure}[b]{0.5\textwidth}
\centering
\includegraphics[width=0.32\linewidth]{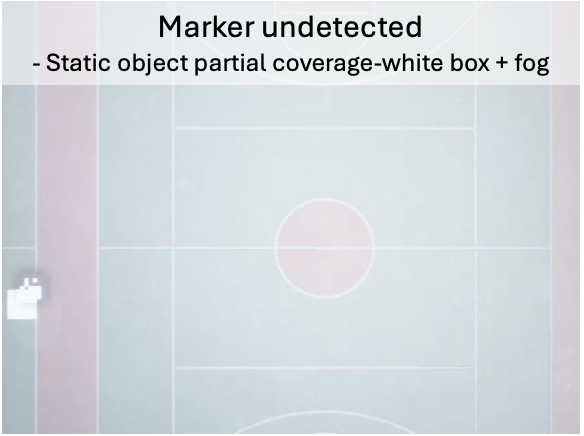}
\includegraphics[width=0.32\linewidth]{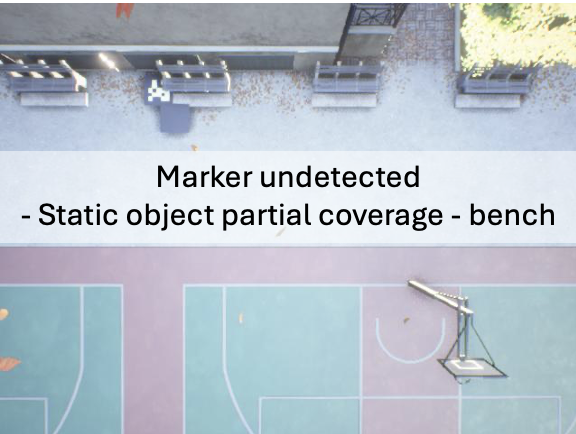}
\includegraphics[width=0.32\linewidth]{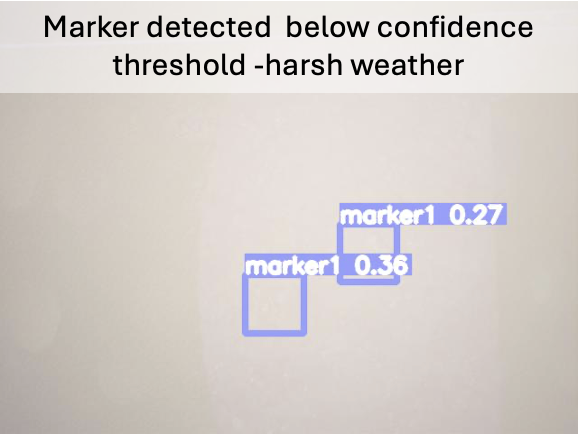}
\caption{Examples of Violation Type II: No marker is detected}
\label{example2}
\end{subfigure}

\begin{subfigure}[b]{0.5\textwidth}
\centering
\includegraphics[width=0.24\linewidth]{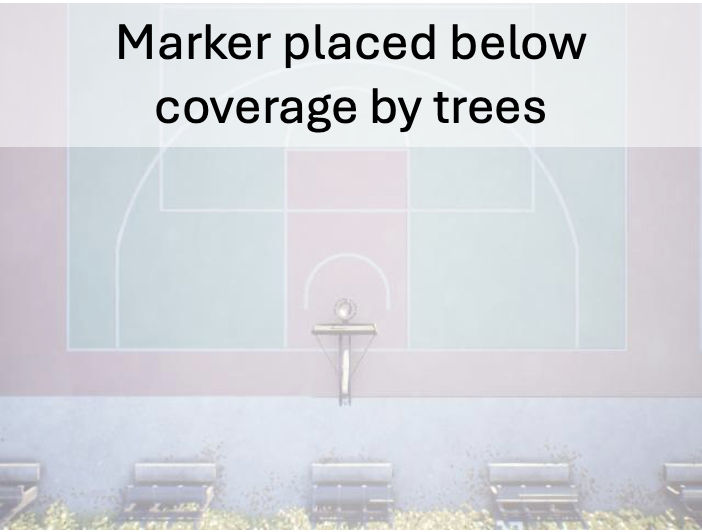}
\includegraphics[width=0.24\linewidth]{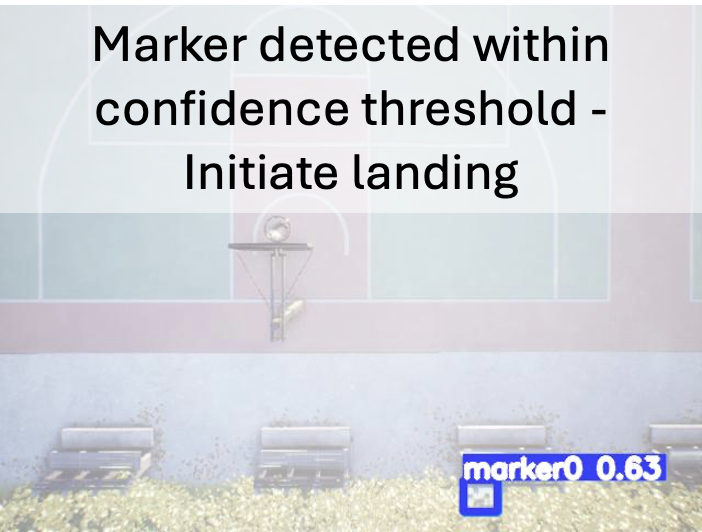}
\includegraphics[width=0.24\linewidth]{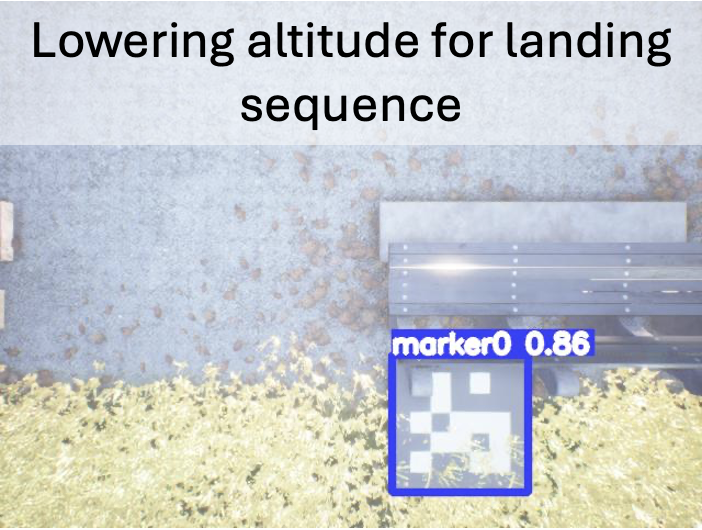}
\includegraphics[width=0.24\linewidth]{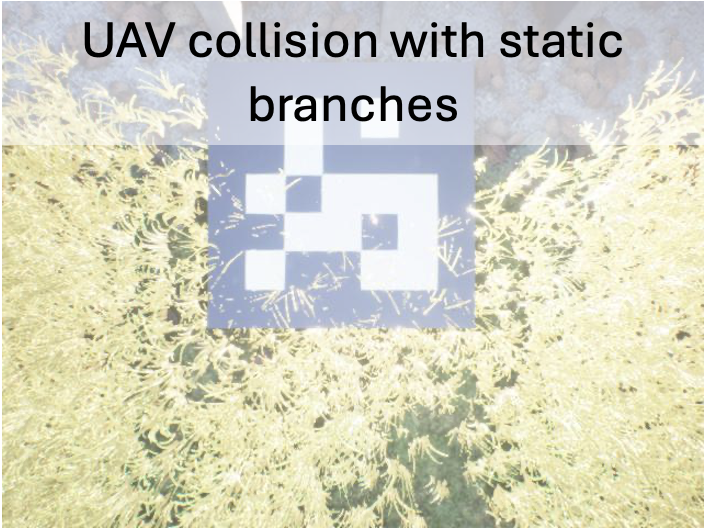}
\caption{Examples of Violation Type III: Static Object Collision}
\label{example3}
\end{subfigure}

% \caption{Examples of violation from simulation (Part 1)}
\label{Examplessim1}

\begin{subfigure}[b]{0.5\textwidth}
\centering
\includegraphics[width=0.49\linewidth]{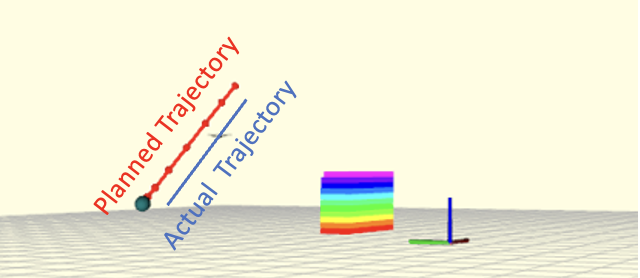}
\includegraphics[width=0.49\linewidth]{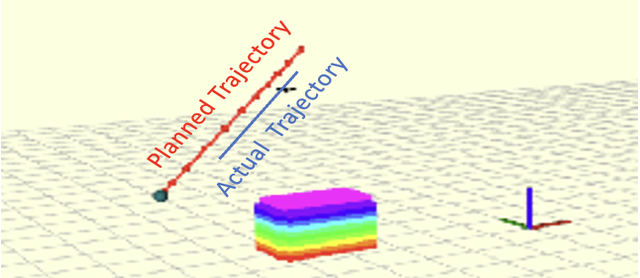}
\caption{Examples of trajectory difference visualized from Rviz}
\label{trajectory difference}
\end{subfigure}

\begin{subfigure}[b]{0.5\textwidth}
\centering
\includegraphics[width=0.24\linewidth]{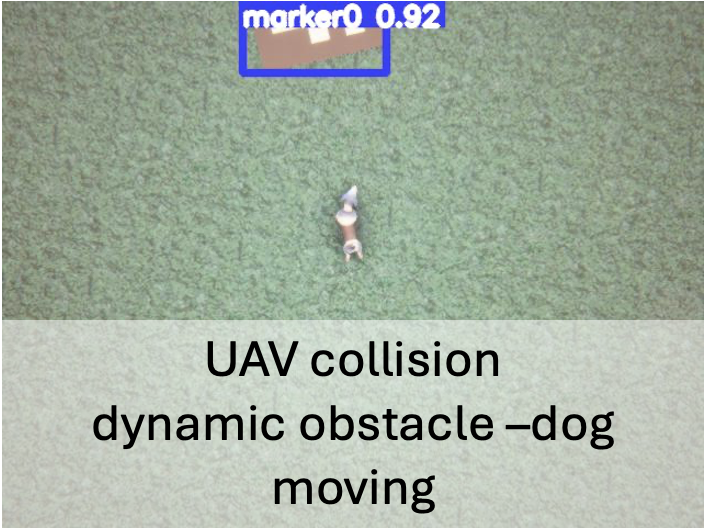}
\includegraphics[width=0.24\linewidth]{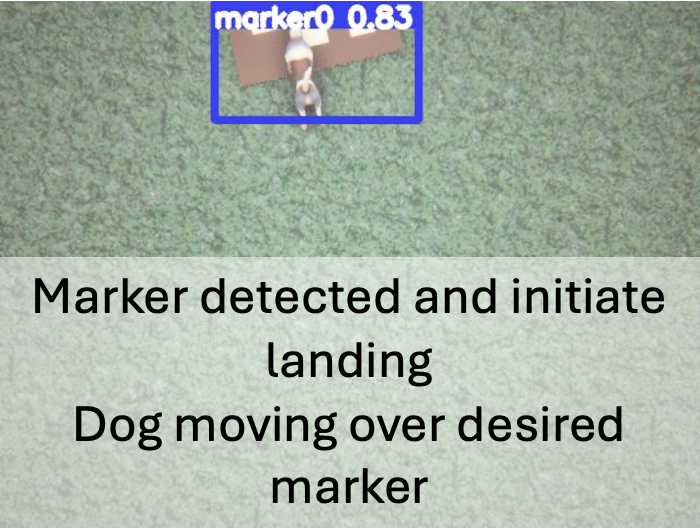}
\includegraphics[width=0.24\linewidth]{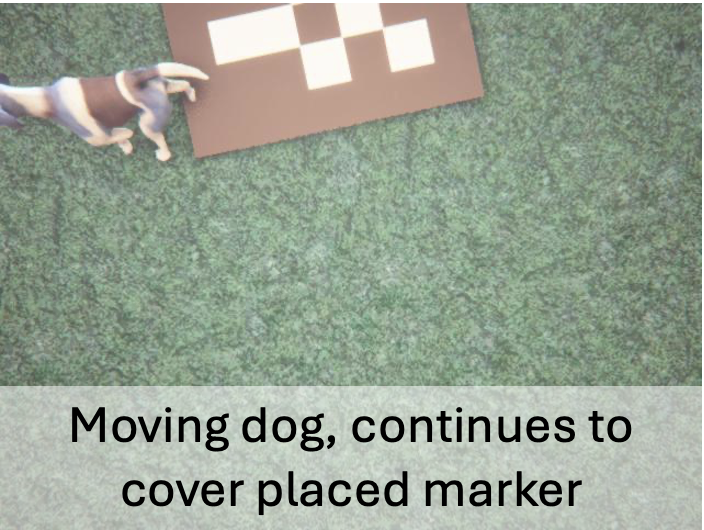}
\includegraphics[width=0.24\linewidth]{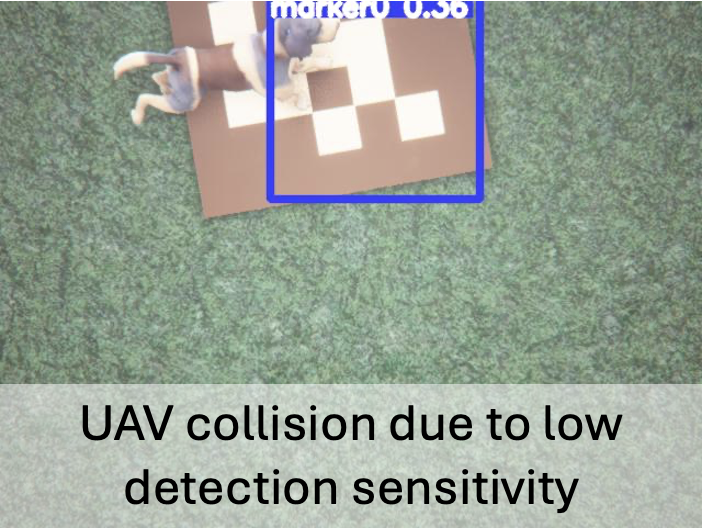}
\caption{Examples of Violation Type IV: Dynamic Object Collision}
\label{example4}
\end{subfigure}

\begin{subfigure}[b]{0.5\textwidth}
\centering
\includegraphics[width=0.24\linewidth]{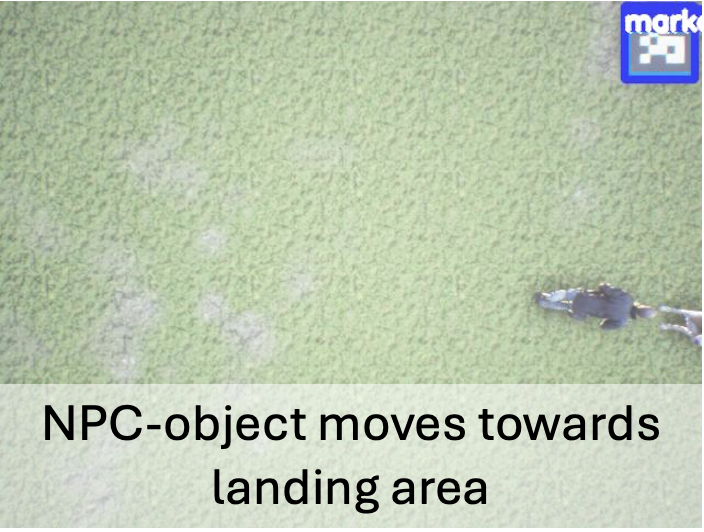}
\includegraphics[width=0.24\linewidth]{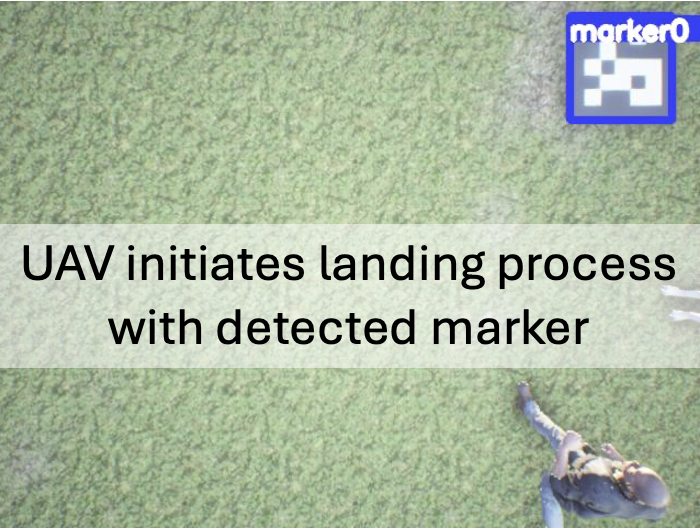}
\includegraphics[width=0.24\linewidth]{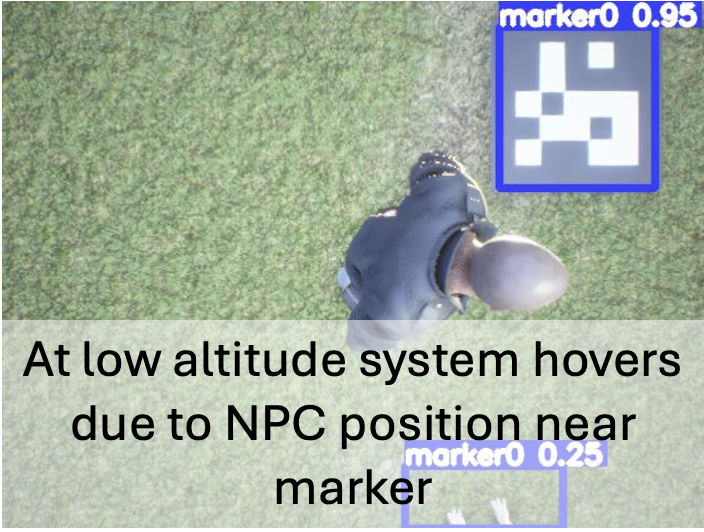}
\includegraphics[width=0.24\linewidth]{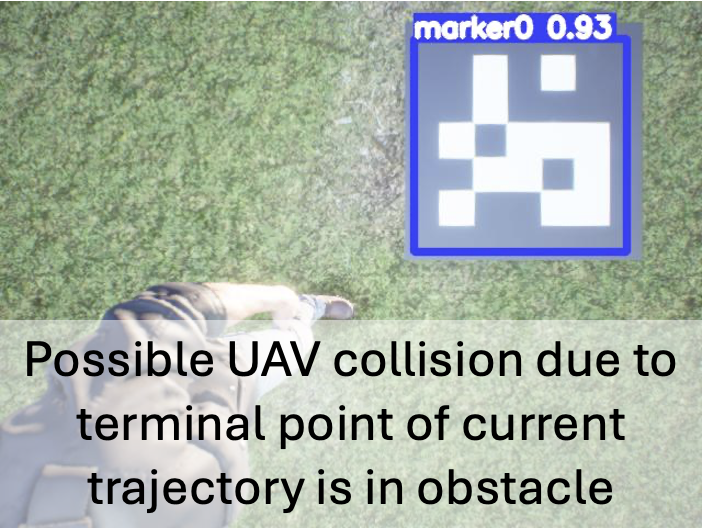}
\caption{Examples of Violation Type V: Planner Crash}
\label{systembug}
\end{subfigure}

\begin{subfigure}[b]{0.5\textwidth}
\centering
\includegraphics[width=0.93\linewidth]{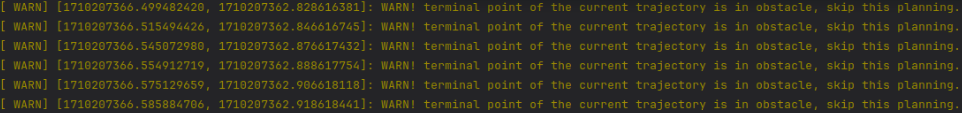}
\caption{Example of crash message}
\label{crash message}
\end{subfigure}

\begin{subfigure}[b]{0.5\textwidth}
\centering
\includegraphics[width=0.45\linewidth]{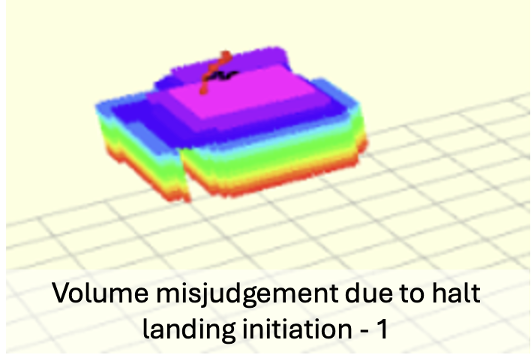}
\includegraphics[width=0.45\linewidth]{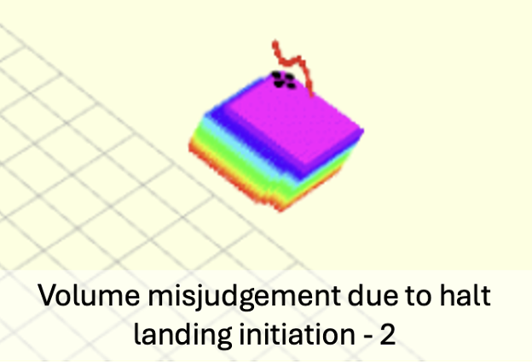}
\caption{Examples of the UAV stack in obstacle visualized from Rviz}
\label{uavinob}
\end{subfigure}

\caption{Examples of violations from simulation}
\label{Examplessim2}
\end{figure}

We classified the violations observed in simulations into specific categories based on a manual analysis of recorded simulations:
%We categorized all the violations identified during simulations into specific types. This classification was based on an analysis of recordings captured during the simulation manually:
%and we first list some example violation types that can be found by most baselines:

\subsubsection{Type I} The first violation type is incorrect landing due to false positive marker detection. Figure \ref{example1} depicts two sequences of landing at incorrect marker locations. The first sequence shows the UAV initiating landing on a ground section erroneously identified as a marker, likely due to ground reflection. Even after losing track of the marker, the UAV proceeds to land incorrectly. The second sequence demonstrates the UAV mistaking a white marker (marker 3 instead of the intended marker 0) as the landing spot and, despite fluctuating confidence, lands there mistakenly.

\subsubsection{Type II} The second type of violation is a failure to land because of a false negative detection. Figure \ref{example2} showcases several instances where no marker is detected. In the first provided example, a white static object partially covers the marker, and the day is foggy. These conditions result in the marker not being detected. In the second example, the marker is placed near a bench%chair
, and the presence of a nearby black object prevents the marker from being detected. In the third example, the perception module has low confidence in the marker location due to the extreme weather conditions. Consequently, no landing occurs.

% \begin{figure}
% \centering

% {\includegraphics[width=0.32\linewidth]{img/no_land/nolanding340_9.jpeg}}
% {\includegraphics[width=0.32\linewidth]{img/no_land/nolanding34_15.jpeg}}
% {\includegraphics[width=0.32\linewidth]{img/no_land/nolanding69_18.jpeg}}

% \caption{Examples of Violation Type II: No marker is detected}

% \label{example2}
% \end{figure}

\subsubsection{Type III} \label{sec: type_3}
The third violation, UAV collision with static objects, occurs when the UAV, detecting a marker and initiating landing, fails to evade nearby branches, resulting in a collision (Figure \ref{example3}). Rviz visualization (Figure \ref{trajectory difference}) shows minor discrepancies between planned and actual trajectories, indicating that slight deviations can lead to collisions near markers surrounded by branches.

\subsubsection{Type IV} 
The fourth violation, involving UAV collisions with moving objects, is illustrated in Figure \ref{example4}, where a UAV does not alter its descent path in response to a dog approaching the marker, resulting in a collision. Analysis indicates that moving ground-level objects, such as the dog, minimally affect the depth data visualized by Rviz, pointing to a low detection sensitivity for objects at low heights. This issue likely stems from the ego-planner's limited capability in processing depth information, a theory supported by real-world experiments.

\subsubsection{Type V} 
The fifth violation, a system-level crash, occurs when a dynamic object moves under a landing UAV, causing the planner to halt landing due to ``the terminal point of current trajectory is in obstacle'' error (Figures \ref{systembug}, \ref{crash message}). Analysis indicates that proximity to moving obstacles can lead to volume misjudgment by the ego-planner, risking a crash (Figure \ref{uavinob}), a finding confirmed by real-world experiments. The detailed distribution of violation types across three landing systems in two maps can be found in the supplementary material.

\subsection{Comparison to Baselines (RQ2)}

\begin{table}
\centering
\caption{Quantitative result of violation case generating efficiency and diversity for different methods, best is marked in bold.}
\resizebox{\columnwidth}{!}{
\begin{tabular}{l l l l l}
\hline
\textbf{Method} & \textbf{Metric} & \textbf{Court} & \textbf{Lawn} \\
\hline
\multirow{5}{*}{\textit{\tool}} & Landing violation \% & \textbf{20.60}\% &\textbf{17.11}\% \\
 & Top-10 & \textbf{42} & \textbf{76} \\
 & Parameter distance & \textbf{0.19} & \textbf{0.19} \\
 & 3D trajectory coverage\% & \textbf{11.24\%} & \textbf{11.94\%} \\
 & Time Consumption (hours) & 12 & 12 \\
\hline
\multirow{5}{*}{\textit{Multi-Obj GA}} & Landing violation \% & 14.25\% & 9.23\% \\
 & Top-10 & 73 & 113 \\
 & Parameter distance & 0.16 & 0.16 \\
 & 3D trajectory coverage & 4.92\% &8.43\% \\
 & Time Consumption (hours) & 11 & 11 \\
\hline
\multirow{5}{*}{\textit{Random}} & Landing violation \% & 9.37\% & 8.52\% \\
 & Top-10 & 205 & 112 \\
 & Parameter distance & 0.13 & 0.13 \\
 & 3D trajectory coverage & 3.51\% & 4.92\% \\
 & Time Consumption (hours) & 11 & 11 \\
\hline
\multirow{5}{*}{\textit{Offline RL Fuzzer}} & Landing violation \% & 2.25\% & 2.13\% \\
 & Top-10 & cannot find & cannot find \\
 & Parameter distance & 0.12 & 0.12 \\
 & 3D trajectory coverage & 1.41\% & 3.51\% \\
 & Time Consumption (hours) & 11 & 11 \\
\hline
\multirow{5}{*}{\textit{Online RL}} & Landing violation \% & 12.75\% & 5.94\% \\
 & Top-10 & 104 & 141 \\
 & Parameter distance & 0.13 & 0.13 \\
 & 3D trajectory coverage & 4.92\% & 7.03\% \\
 & Time Consumption (hours) & 11 & 11 \\
\hline
\multirow{5}{*}{\textit{Surrogate trained RL with random scenario}} & Landing violation \% & 14.19\% & 13.53\% \\
 & Top-10 & 67 & 108 \\
 & Parameter distance & 0.13 & 0.13 \\
 & 3D trajectory coverage & 8.43\% & 8.43\% \\
 & Time Consumption (hours) & 12 & 12 \\
\hline
\end{tabular}}
\label{tab:exp2_quantitative_single_column}
% \vspace{-0.8cm}
\end{table}

As shown in Table \ref{tab:exp2_quantitative_single_column}, \tool\ achieved the best performance on both maps across all metrics. On the \texttt{Court} map, \tool\ found 20.60\% more violations, up to 18.35\% better than SOTA methods in 400 runs. It excelled in diversity, with the highest Parameter distance (0.19) and 3D-Trajectory coverage (0.16\%). \tool\ also required only 42 test cases to reveal the \textbf{Top-10} bugs. Similar results were observed for the \texttt{Lawn} map.
%As shown in Table \ref{tab:exp2_quantitative_single_column}, \tool \ achieved the best performance on both maps for all metrics when compared to baselines.
%In the \texttt{Court} map, \tool\ was able to find more violations with 20.60\% which is up to 18.35\% better than SOTA methods in 400 runs. In evaluating diversity, \tool\ excelled, achieving the highest Parameter distance at 0.19 and leading in 3D-Trajectory coverage with 0.16\% exploration on the whole map. These results highlight its exceptional ability to discover diverse violation cases. In terms of efficiency of finding bugs, \tool\ performs best with 42 test cases to reveal the \textbf{Top-10} bugs. Similar results were obtained for the \texttt{Lawn} map. 
%\jz{one question reviewers might ask is how many test cases you run per method, give the number here and it would be great if the number is identical}. 
%$RLaGA$ was also able to find all violation types in $64$ rounds while the other baselines were either unable to discover all types or took more rounds to find this diversity (132 rounds for Diverse GA).
Moreover, GARL has a similar time consumption compared to other baselines. Since all testing methods require running the test in the simulation while the RL agent runs side-by-side with the Airsim simulator on the same machine and calls the Airsim API, the running time overhead is mainly due to collision avoidance planning and the longer trajectory of flight caused by the complex interplay with the trajectory of dynamic objects. The overall time difference between methods is not significant. For example, running 400 test budgets for GARL and surrogate-trained RL with random scenarios takes around 12 hours on an RTX 3090 GPU, compared to 11 hours for other baselines. However, our method requires an additional 12 hours for training the RL agent in the surrogate environment, which is only needed once for all the maps. For the computational resources necessary for RL, we use a two-layer MLP, which requires just 2GB of GPU memory. We include a more detailed statistical analysis of all baselines in the supplementary material.

\par 
For \textit{Multi-Objective GA}, performance on the \texttt{Lawn} map is only slightly better than the \textit{Random} baseline. This is because GA often passes on configurations causing violations. Most violations in \texttt{Lawn} are due to collisions, which \textit{Multi-Objective GA} struggles to detect. The \textit{Surrogate trained RL with random scenario} baseline performs closely to \tool, highlighting the importance of online RL.

%To further analyse the performance of other baselines, such as \textit{Multi-Objective GA}, performance on the \texttt{Lawn} map is only slightly better to \textit{Random} baseline. This outcome stems from GA's tendency to pass on environment configurations that cause violations to subsequent generations. However, as analyzed in RQ1, the majority of violations in the \texttt{Lawn} map result from collisions, a type of violation that \textit{Multi-Objective GA} struggles to find effectively by predefined dynamic object's trajectory. Observing the \textit{Surrogate trained RL with random sceanrio} baseline on the same map, its performance closely follows \tool, suggesting the importance of online RL. 

% Additionally, \textit{Multi-Objective GA} and \textit{Online RL-2} baselines can also serve as the ablation study of \tool. From the result, we can see that .

% \begin{figure}
% \centering

% \includegraphics[width=0.8\linewidth]{img/reward_curve.jpg}
   
% \caption{\lf{Comparison of Reward for Different Learning Space in Surrogate Training}}
% \vspace{-0.5cm}
% \label{reward_training}
% \end{figure}

\par 
Both \textit{Online RL} and \textit{Offline RL Fuzzer} underperform. GA effectively reduces the learning space for RL training, with the GA-reduced RL agent converging in 800 episodes, compared to 4200 episodes for the full learning space. This highlights GA's advantage in facilitating efficient RL training. RL agents struggle to converge in complex real-world spaces without GA's targeted reduction. Detailed rewards figures are in the supplementary document.

\par 
\tool\ and the \textit{Surrogate trained RL with random scenario} uniquely detected \textit{Type IV} and \textit{Type V} violations. However, without GA, the latter struggles in complex scenarios like the \texttt{Court} map. GA aids by identifying potential seeds for violation detection, simplifying RL's requirements, and highlighting promising candidates, especially in complex scenarios.

\subsection{Real-World Reproduction (RQ3)}
\label{re}

\par
\begin{figure}
\centering

\begin{subfigure}[b]{0.23\textwidth}
\centering
    \includegraphics[width=0.9\textwidth]{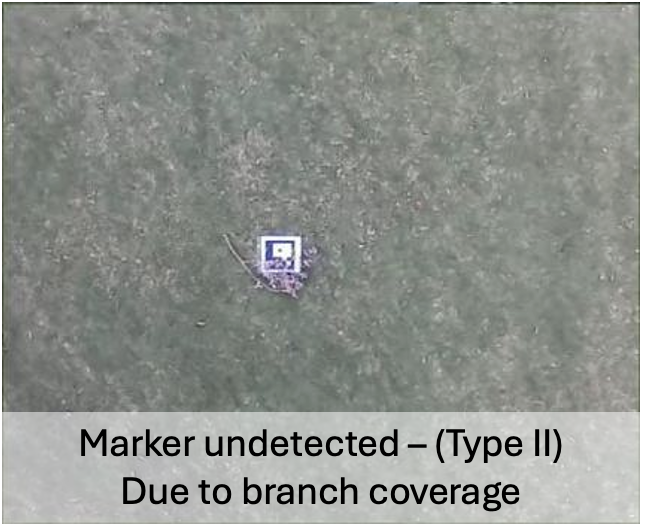}    
    \caption{False negative detection}
    \label{fndetection}
\end{subfigure}
\begin{subfigure}[b]{0.23\textwidth}
\centering
    \includegraphics[width=0.9\textwidth]{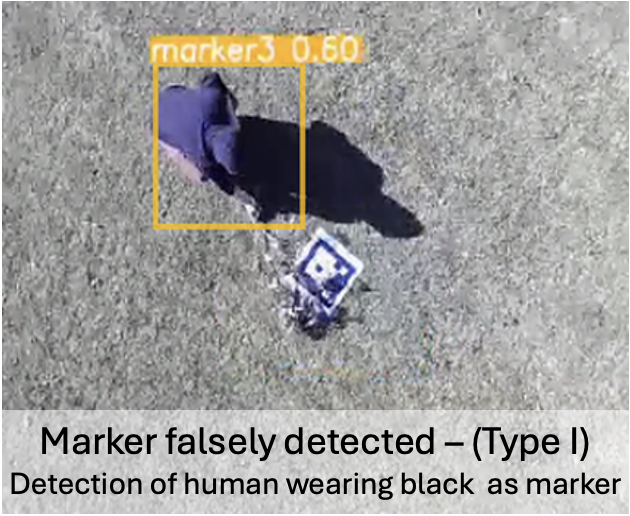}    
    \caption{False positive detection}
    \label{fpdetection}
\end{subfigure}

\begin{subfigure}[b]{0.48\textwidth}
\centering
{\includegraphics[width=0.4\linewidth]{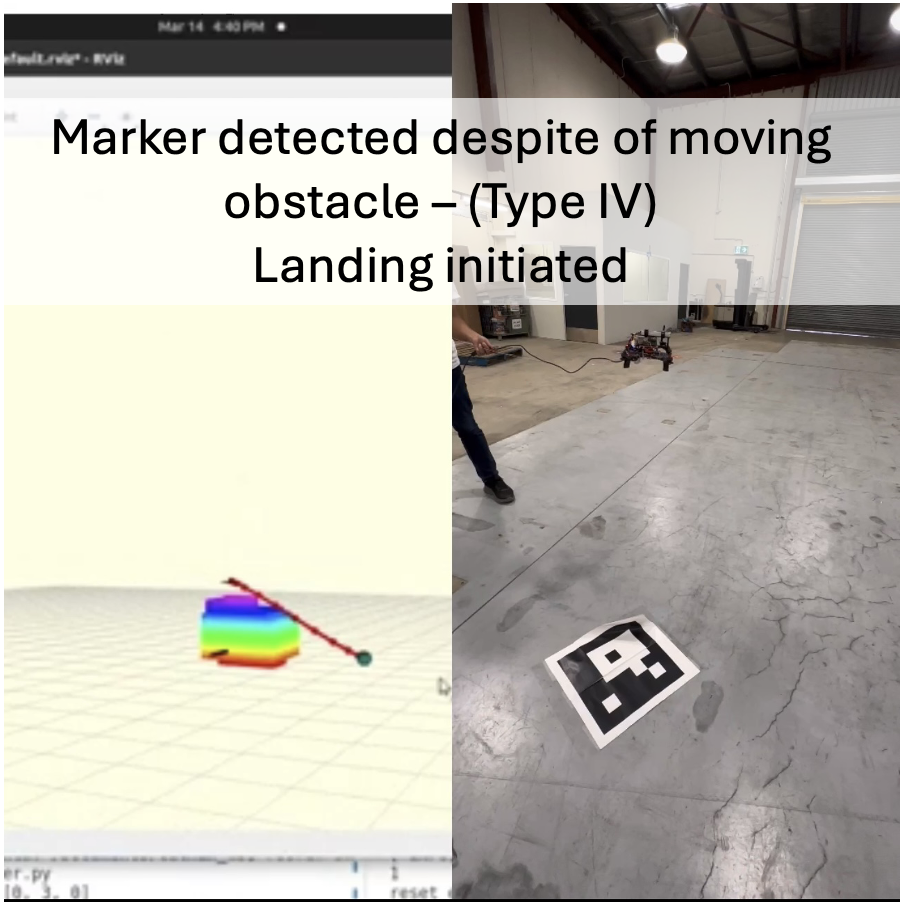}}
{\includegraphics[width=0.4\linewidth]{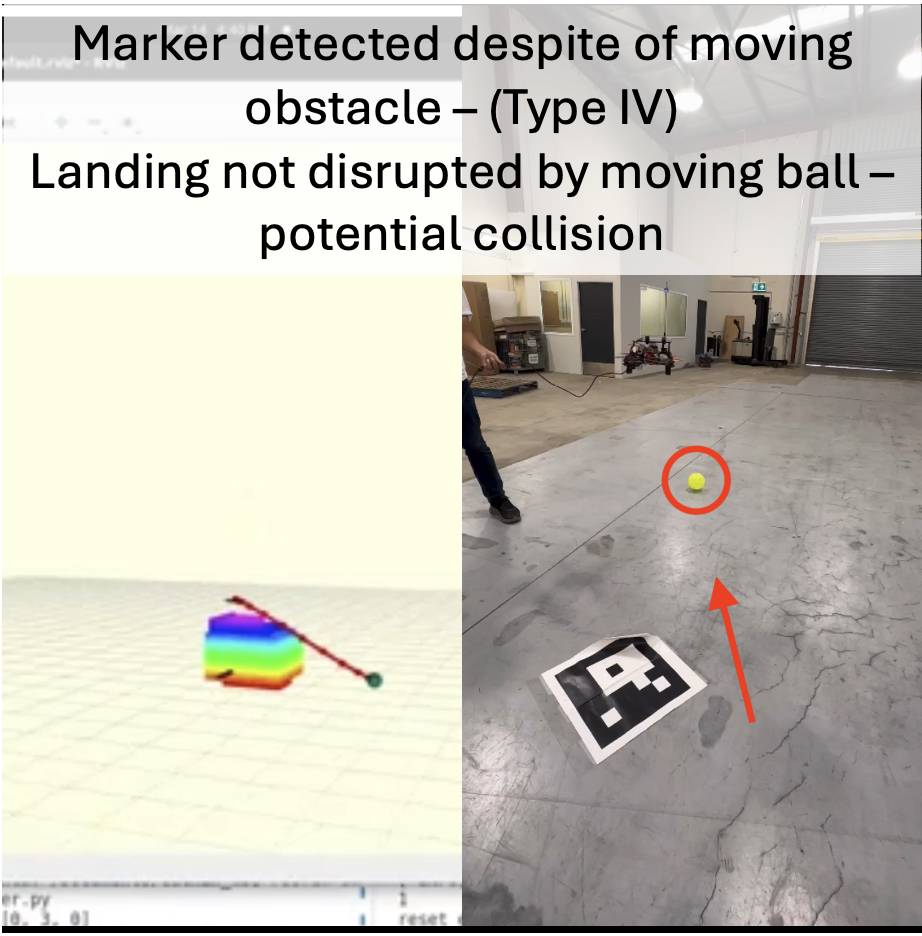}}
\caption{Real-world experiment for \textit{Type IV}}
        \label{rwc2}
\end{subfigure}
\begin{subfigure}[b]{0.48\textwidth}
\centering
{\includegraphics[width=0.4\linewidth]{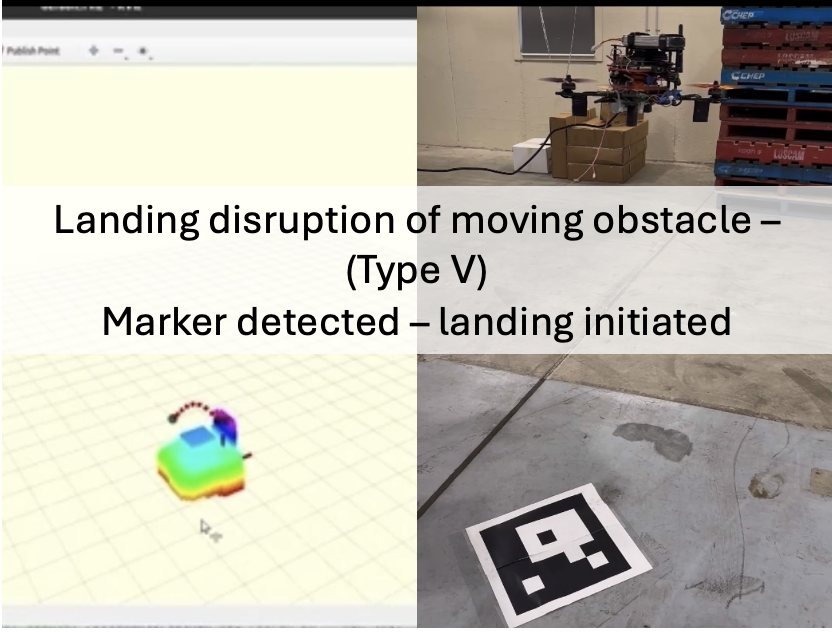}}
{\includegraphics[width=0.4\linewidth]{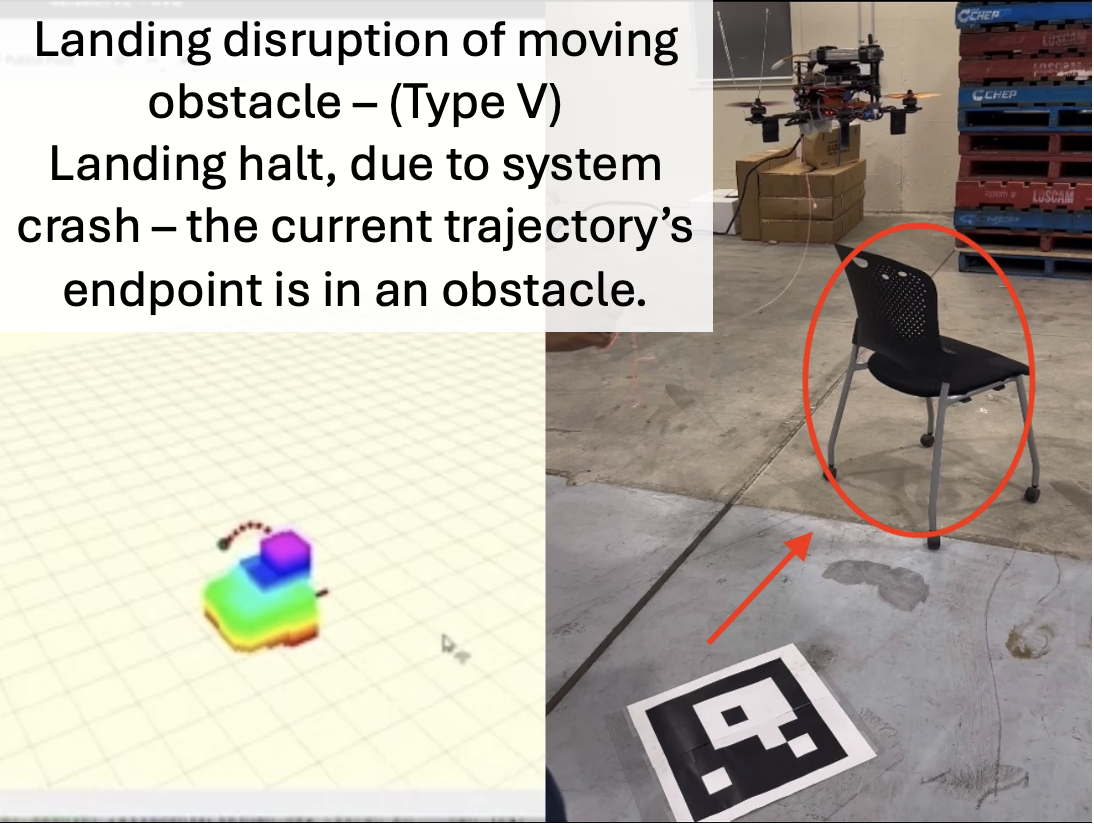}}
\caption{Real-world experiment for \textit{Type V}}
        \label{rwc1}
\end{subfigure}

\caption{Example of real-world violations reproduction}
% \vspace{-0.8cm}
\label{rwdection}
\end{figure}

 We validated the identified violation cases %identified by \tool\ 
in simulations with real-world tests. We reproduced \textit{Type I} and \textit{Type II} violations on a lakeside lawn under sunny conditions, a constrained physical environment. The test used a $0.75m \times 0.75m$ marker and a UAV flying at a height of $5m$, scaled to 50\% of the simulation setting. Figure \ref{fndetection} shows a \textit{Type II} false negative detection where the marker is initially identified but becomes undetectable when partially obscured by a branch. Figure \ref{fpdetection} shows a \textit{Type I} false positive detection where a person wearing a black hoodie is detected as the ArUco marker.

%We validated the violation cases identified by \tool\ in simulations with real-world tests. We first reproduced \textit{Type I} and \textit{Type II} violations on a lakeside lawn under sunny conditions which is a constrained physical environment. The test involved a marker of size $0.75m \times 0.75m$ and a UAV flying at a height of $5m$, proportional scaling to 50\% of the setting in the simulation environment. Figure \ref{fndetection} shows a false negative detection (\textit{Type II}) in the real world where initially, the marker is correctly identified. However, once partially obscured by a branch, it becomes undetectable. Figure \ref{fpdetection} indicates a false positive detection (\textit{Type I}) in the real world, where a person wearing a black hoodie is detected as the ArUco marker.

% The test involved a marker of size $1m \times 1m$ and a UAV flying at a height of $10m$, which matches the setting in the simulation. 

%, and a man in a black jacket as the dynamic object.

% \begin{figure}
% \centering
% \includegraphics[width=0.5\textwidth]{img/real_world_test.pdf}
% \end{figure}

\par For \textit{Type III} violations, we did not reproduce the scenario in the real world due to safety concerns. Validating this type would require allowing the UAV to execute its planned trajectory, which was deemed infeasible. The real-world performance of a UAV with only GPS input typically has a position estimate accurate within 1 meter \cite{fornasierequivariant}. However, GPS signal loss can cause larger deviations \cite{pang2020uav}. The identified violation underscores the need for more robust UAV landing solutions and demonstrates our framework's capability to help developers identify subtle issues earlier in the development lifecycle.

%For \textit{Type III} violations, we refrained from reproducing the scenario in the real world because validating this type would require allowing the UAV to execute its planned trajectory. However, due to safety concerns, permitting the UAV to follow through with the trajectory was deemed infeasible. The real-world performance of a UAV with only GPS input has a position estimate that is typically accurate within than 1 meter of the truth position \cite{fornasierequivariant}. In addition, because of the loss of GPS signal, the deviation may be larger as analyzed in the work~\cite{pang2020uav}. The founded violation type indicates that it is important to develop more robust solutions for UAV landing. It also demonstrates the capability of our work to help developers find subtle issues earlier in the development life cycle.  

% \par For violation types \textit{IV} and \textit{V}, we conducted controlled real-world tests due to safety considerations. We began by analyzing potential causes for these collision-related violations. 
\par For \textit{Type IV} violations, as shown in Figure \ref{rwc2}, a real-world scenario was observed where a ball moving across the depth camera's field of view did not alter the captured depth information, potentially leading to a collision during the UAV's landing phase. This issue aligns with our simulation experiment, where near-ground objects like dogs are not detected as obstacles and cause collisions, as shown in Figure~\ref{example4}.

%For \textit{Type IV} violations, as illustrated in Figure \ref{rwc2}, a comparable real-world scenario was observed when a ball moving across the depth camera's field of view did not alter the captured depth information. This oversight could lead to a potential collision during the UAV's landing phase. This problem is in accordance with the experiment in simulation that near-ground objects such as dogs are not detected as an obstacle and cause collision, as shown in Figure~\ref{example4}.

% This underscores the limitations of depth cameras in accurately identifying small, low-lying objects.

\par For \textit{Type V} violations, a quick movement of an object closely beneath the UAV can trigger a crash in the planner. In the real-world test shown in Figure \ref{rwc1}, the planner initially captures depth information accurately. However, when a chair is rapidly moved underneath the UAV, it is perceived as an obstacle, raising the error ``the terminal point of current trajectory is in obstacle'', underscoring the issue's relevance in real scenarios.

\section{Threats to Validity}
\label{TV}

\textbf{Construct Validity}: 
The construct validity is challenged by adapting landing systems and baseline methods for comparison. We developed three auto-landing systems: \textit{OpenCV-MLS}, \textit{TPHYolo-MLS}, and \textit{MM-MLS}. 
Three marker-based landing systems were selected to represent different capability levels and demonstrate iterative evolution through our testing approach. OpenCV-MLS, the only open-source baseline implementation \cite{Goodrobots_Vision_Landing} used by our industry partner, initially struggled with detecting small markers. To improve accuracy and robustness, we integrated the TPHYolo detector, known for detecting small objects in UAV images, creating TPHYolo-MLS. The most robust system, MM-MLS, enhances TPHYolo-MLS by adding the Ego-planner, a state-of-the-art path planner for real-time, collision-free navigation, similar to industrial autonomous driving systems like Apollo and Autoware. This selection illustrates iterative improvements enabled by our testing methodology, with each system representing a progression in functionality, allowing us to evaluate how effectively \tool\ identifies issues and drives enhancements toward a more robust and reliable marker-based landing system.

Due to limited research on testing marker-based landing systems, direct baseline methods are challenging to find. We adopted state-of-the-art search-based~\cite{tian2022mosat}, fuzzer~\cite{bottinger2018deep}, and online~\cite{lu2022learning} test generation methods from the software testing community, aligning their fitness or reward functions with ours. Although we found a method called Adaptive Stress Testing with backward training~\cite{koren2021finding}, we did not compare it because RL with a full learning space fails to converge even in a simplified surrogate environment, making it impractical to obtain expert trajectories for backward training. The training curves can be found in the supplementary material.
%Due to limited research on testing marker-based landing systems, finding direct baseline methods for comparison is challenging. To solve the problem, we adopted the state-of-the-art search-based~\cite{tian2022mosat}, fuzzer \cite{ bottinger2018deep} and online~\cite{lu2022learning} test generation methods in the Software testing community. For those methods, we kept their main ideas and adopted their fitness functions or reward functions aligned with ours. We also found a test generation method called Adaptive Stress Testing with backward training \cite{koren2021finding} but did not compare with this work. The reason is that even within the simplified surrogate environment, RL with a full learning space fails to converge, \lf{the training curves can be found in the supplementary material}. Consequently, it becomes impractical to acquire expert trajectories in scenarios characterized by such extensive learning dimensions for backward training.
\par Another construct validity issue arises from GPS errors. In the simulation environment, we did not account for GPS errors because AirSim provides ground truth GPS data and lacks interfaces for modifying it. However, when setting up markers, we randomly positioned the center point within a 7.5-meter radius of the marker's GPS coordinate. The vision-based landing system is designed to handle inaccurate GPS landing locations. In real-world experiments, we considered GPS as a potential factor for Type III violations (collisions with static objects). To reproduce other violation cases, we tested the system in a stable GPS environment. Additionally, during flights, the Ardupilot controllers use the Extended Kalman Filter (EKF) to manage potential GPS errors \cite{ArduPilot_EKF_Overview}.

%\lf{Another construct validity issue is raised from GPS. In the simulation environment, we did not account for GPS errors at the time of writing the paper because AirSim provides ground truth GPS sensor data and does not offer interfaces for modifying GPS data. However, when setting up markers in each testing scenario, we did not place the marker at the exact GPS coordinate. Instead, we randomly positioned the center point of the marker within a 7.5-meter radius of the GPS coordinate of the marker. Additionally, the vision-based landing system is designed to handle issues with inaccurate GPS landing locations. In real-world experiments, we considered and analyzed GPS as a potential factor causing Type III violations (collision with static objects). To reproduce other types of violation cases, we tested the system in a stable GPS environment. Furthermore, during flights, the underlying Ardupilot controllers utilize the Extended Kalman Filter (EKF) to manage potential GPS errors \cite{ArduPilot_EKF_Overview}.}

% Another threat to the construct validity is our inability to compare \tool with Adaptive Stress Testing that incorporates backward training \cite{koren2021finding}. This limitation arises because, 

\textbf{External Validity}: The primary external validity concern is the generalizability of our method. To tackle this, we tested our search method across various landing systems, including \textit{OpenCV-MLS}, \textit{TPHYolo-MLS}, and \textit{MM-MLS}, and on maps representing real-world deployment sites of our industry partners. Our method successfully identified diverse violation cases across these landing systems and maps.
%The main threat to external validity is regarding the generalization of the proposed method. To address the problem, we have conducted experiments to generalize our search method on different landing systems including \textit{OpenCV-MLS}, \textit{TPHYoLo-MLS}, and \textit {MM-MLS}. In addition, we tested these landing systems on different maps based on real-world deployment sites of our industry partners. Our method has demonstrated the capability to identify diverse violation cases across these different automatic landing systems and maps.

\textbf{Internal Validity}: A potential threat to internal validity is the fidelity of the simulation environment. To address this concern, we chose maps relevant to real-world applications. The marker-based landing system we tested is designed for delivery purposes. Our industry research collaborators target grassy lawns and concrete hospital courts with additional markings and ground obstacles. The lawn map represents rural farms for logistics delivery, while the basketball map represents hospital and industrial zones for urgent components and medical supplies delivery. For each deployment map, we incorporated dynamic objects found in the actual deployment sites into AirSim. We ensured our maps closely resembled real deployment environments. Positive outcomes from our real-world tests further alleviate this validity concern.

%\lf{To address this concern, we first choose the map against the real-world application scenario. The marker-based landing system we tested is designed for delivery purposes. Our industry research collaborators target two types of areas: grassy lawns and concrete basketball courts with additional markings and ground obstacles. The lawn map represents rural farms where logistics need to be delivered, while the basketball map represents hospital and industrial zones, where urgent components and medical supplies need to be delivered.}

%\lf{Then, for each deployment map, we incorporated dynamic-object types found in the actual corresponding deployment sites into Airsim. Additionally, we ensured our maps closely resembled the real deployment environments. The positive outcomes from our real-world tests further alleviate this validity concern.}

\par 
Another potential threat is real-world reproduction. To replicate the simulation results, we adjusted controllable environmental factors and considered uncontrollable ones like dust, fog, and snow. Adjustable factors included marker locations, nearby objects, time of day for lighting changes, and a regulated wind speed of 3 m/s for lakeside tests. For type I (wrongly recognizing markers) and type II (missing true markers) violations, we varied marker locations, placed objects near markers, and conducted tests at different times of the day. We configured the simulation’s camera parameters based on physical camera specifications and aligned the UAVs' flight control parameters to replicate the tested flight patterns. Our simulation platform and testing methods generated detailed logs for fault analysis, as shown in the RQ1 results (P8). We successfully reproduced these violations in real-world settings based on identified root causes. However, we faced limitations not present in the simulation, such as minor lens stains, blurry images, excessive glare, and poor network connections affecting detection models. These issues could trigger violations in real-world testing. We plan to conduct extensive field testing with industry partners in diverse environments to gather feedback. This feedback will help build a more high-fidelity co-simulation or hardware-in-the-loop simulation platform with Simulink.

\section{Discussion}
Our approach effectively detects issues in path planning, which is crucial for UAV safe landing, monitoring, surveying, and other autonomous systems, including vehicles. This method can be applied early in the software development life cycle, during the prototype and design stages, to detect bugs and guide design choices for machine learning models and system architecture. Early detection helps avoid costly fixes post-deployment. Using our simulation platform, we detected bugs and improved landing solutions, resulting in three generations of auto-landing systems with different machine learning models and architectures. Our method also aids in business operation risk assessment to identify potential risks and determine safe deployment areas, and detect early bugs in other critical components, such as the perception module.

Using RL for online testing is challenging due to difficulties in training agents to converge in complex action spaces. \tool\ allows practitioners to conduct a one-time surrogate training period of approximately 12 hours for our landing system action spaces, after which the trained agent can be applied across various maps and deployment environments in simulation. This method is effective in detecting subtle and complex issues, particularly in path planning, a crucial component in many autonomous systems. Early detection during the prototype and design stages is cost-effective and can significantly influence design decisions. For example, we successfully used this approach to modify key perception models, path planning, and architecture in three landing systems.

\section{Conclusion}
\label{conclusion}
This paper introduces a method for generating diverse violation cases in marker-based UAV landing systems. Our approach integrates GA to reduce the RL search space and a low-fidelity proxy to improve RL convergence. Experiments show that \tool\ detects 18.35\% more violations with 58\% greater diversity than existing methods, uncovering previously undetected violation types validated in real-world tests. These findings highlight \tool's efficacy in early error detection, enhancing safety and reducing costs. Future work includes extending the RL algorithm to multi-agent RL for collaborative behaviors, broadening test scenarios, and applying \tool\ to learning-enabled autonomous systems like vehicles and humanoid robots.

\section*{Acknowledgement}
This work is supported by Australia Research Council grants LP210100337, LP190100676, and DP210102447.

\newpage

% used the genetic algorithm as both the global searching method and the local searching method to generate more delicate violation cases.

%
% ---- Bibliography ----
%
% BibTeX users should specify bibliography style 'splncs04'.
% References will then be sorted and formatted in the correct style.
%
\bibliographystyle{splncs04}
\bibliography{reference}
%
% \begin{thebibliography}{}
% \bibitem{ref_article1}
% Author, F.: Article title. Journal \textbf{2}(5), 99--110 (2016)

% \bibitem{ref_lncs1}
% Author, F., Author, S.: Title of a proceedings paper. In: Editor,
% F., Editor, S. (eds.) CONFERENCE 2016, LNCS, vol. 9999, pp. 1--13.
% Springer, Heidelberg (2016). \doi{10.10007/1234567890}                                                                            

% \bibitem{ref_book1}
% Author, F., Author, S., Author, T.: Book title. 2nd edn. Publisher,
% Location (1999)

% \bibitem{ref_proc1}
% Author, A.-B.: Contribution title. In: 9th International Proceedings
% on Proceedings, pp. 1--2. Publisher, Location (2010)

% \bibitem{ref_url1}
% LNCS Homepage, \url{http://www.springer.com/lncs}. Last accessed 4
% Oct 2017
% \end{thebibliography}
\end{document}